\documentclass{elsarticle}
\usepackage{txfonts}
\usepackage{epsfig}
\usepackage{subfigure}

\makeatletter
\renewcommand{\paragraph}{\@startsection{paragraph}{4}{0ex}%
   {-3.25ex plus -1ex minus -0.2ex}%
   {1.5ex plus 0.2ex}%
   {\normalfont\normalsize\itshape}}
\makeatother

\stepcounter{secnumdepth}
\stepcounter{tocdepth}

\begin{document}

\begin{frontmatter}

\title{Characterization of a tagged $\gamma$-ray beam line at the DA$\Phi$NE Beam Test Facility }

\author[label7]{P.W. Cattaneo \corref{cor1}} \cortext[cor1]{Corresponding author}
        \ead{Paolo.Cattaneo@pv.infn.it }
\author[label1]{A. Argan}
\author[label7]{F. Boffelli}
\author[label5]{A. Bulgarelli}
\author[label21]{B. Buonomo}
\author[label3,label4]{A.W. Chen}
\author[label22]{F. D'Ammando}
\author[label4]{L. Foggetta \corref{cor2}} \cortext[cor2]{Now at
LAL-CNRS, F-91898 Orsay, France} 
\author[label2,label4]{T. Froysland}
\author[label5]{F. Fuschino}
\author[label8]{M. Galli}
\author[label5]{F. Gianotti}
\author[label3]{A. Giuliani} 
\author[label6]{F. Longo}
\author[label5]{M. Marisaldi }
\author[label21]{G. Mazzitelli}
\author[label18]{A. Pellizzoni} 
\author[label12]{M. Prest}
\author[label13]{G. Pucella}
\author[label21]{L. Quintieri}
\author[label7]{A. Rappoldi}
\author[label1,label2]{M. Tavani} 
\author[label5]{M. Trifoglio }
\author[label18]{A. Trois }
\author[label9]{P. Valente}
\author[label6]{E. Vallazza }
\author[label19]{S. Vercellone} 
\author[label3]{A. Zambra }
\author[label6]{G. Barbiellini}
\author[label3]{P. Caraveo} 
\author[label1]{V. Cocco}
\author[label1]{E. Costa}
\author[label1]{G. De Paris}
\author[label1]{E. Del Monte}
\author[label5]{G. Di Cocco}
\author[label1]{I. Donnarumma}
\author[label1]{Y. Evangelista} 
\author[label1]{M. Feroci}
\author[label4,label17]{A. Ferrari} 
\author[label3]{M. Fiorini}
\author[label5]{C. Labanti}
\author[label1]{I. Lapshov}
\author[label1]{F. Lazzarotto}
\author[label9]{P. Lipari}
\author[label10]{M. Mastropietro}
\author[label3]{S. Mereghetti}
\author[label5]{E. Morelli}
\author[label6]{E. Moretti}
\author[label11]{A. Morselli}
\author[label1]{L. Pacciani}
\author[label3]{F. Perotti} 
\author[label1,label2,label11]{G. Piano}
\author[label2,label11] {P. Picozza}
\author[label12]{M. Pilia}
\author[label1]{G. Porrovecchio}
\author[label13]{M. Rapisarda }
\author[label1]{A. Rubini} 
\author[label1,label2]{S. Sabatini }
\author[label1]{P. Soffitta}
\author[label2,label11]{E. Striani }
\author[label1,label2]{V. Vittorini }
\author[label9]{D. Zanello }
\author[label14]{S. Colafrancesco} 
\author[label14]{P. Giommi}
\author[label14]{C. Pittori} 
\author[label14]{P. Santolamazza }
\author[label14]{F. Verrecchia} 
\author[label15]{L. Salotti }

\address[label1] {INAF/IASF-Roma, I-00133 Roma, Italy}
\address[label2] {Dip. di Fisica, Univ. Tor Vergata, I-00133 Roma,Italy}  
\address[label3] {INAF/IASF-Milano, I-20133 Milano, Italy}
\address[label4] {CIFS-Torino, I-10133 Torino, Italy}
\address[label5] {INAF/IASF-Bologna, I-40129 Bologna, Italy}
\address[label6] {INFN Trieste, I-34127 Trieste, Italy} 
\address[label7] {INFN-Pavia, I-27100 Pavia, Italy}
\address[label8] {ENEA-Bologna, I-40129 Bologna, Italy}
\address[label9] {INFN-Roma La Sapienza, I-00185 Roma, Italy}
\address[label10] {CNR-IMIP, Roma, Italy}
\address[label11] {INFN Roma Tor Vergata, I-00133 Roma, Italy} 
\address[label12] {Dip. di Fisica, Univ. Dell'Insubria, I-22100 Como, Italy}
\address[label13] {ENEA Frascati,  I-00044 Frascati (Roma), Italy}
\address[label14] {ASI Science Data Center, I-00044 Frascati(Roma), Italy} 
\address[label15] {Agenzia Spaziale Italiana, I-00198 Roma, Italy} 
\address[label16] {Osservatorio Astronomico di Trieste, Trieste, Italy} 
\address[label17] {Dip. Fisica, Universit\'a di Torino, Torino, Italy}
\address[label18] {INAF-Osservatorio Astronomico di Cagliari, 
localita' Poggio dei Pini, strada 54, I-09012 Capoterra, Italy} 
\address[label19] {INAF-IASF Palermo, Via Ugo La Malfa 153, I-90146  Palermo, Italy}
\address[label20] {Dip. Fisica Univ. di Trieste, I-34127 Trieste, Italy} 
\address[label21] {INFN Lab. Naz. di Frascati, I-00044 Frascati(Roma), Italy} 
\address[label22] {INAF-IRA Bologna, Via Gobetti 101, I-40129 Bologna, Italy}

\date{\Today}

\begin{abstract}

At the core of the AGILE scientific instrument, designed to operate on a satellite,
there is the Gamma Ray Imaging Detector (GRID) consisting of a Silicon Tracker (ST), a 
Cesium Iodide Mini-Calorimeter and an Anti-Coincidence system  
of plastic scintillator bars.
The ST needs an on-ground calibration with a $\gamma$-ray beam to validate the simulation used 
to calculate the energy response function and the effective area versus the energy and the 
direction of the $\gamma$ rays.
A tagged $\gamma$-ray beam line was designed at the Beam Test Facility (BTF) of the INFN Laboratori
Nazionali of Frascati (LNF), based on an electron beam generating $\gamma$ rays through 
bremsstrahlung in a position-sensitive target. The $\gamma$-ray energy is deduced by 
difference with the post-bremsstrahlung electron energy \cite{prest}-\cite{hasan}.  
The electron energy is measured by a spectrometer consisting of a dipole magnet and 
an array of position sensitive silicon strip detectors, the Photon Tagging System (PTS).
The use of the combined BTF-PTS system as tagged photon beam requires understanding the 
efficiency of $\gamma$-ray tagging, the probability of fake tagging, the energy resolution and 
the relation of the PTS hit position versus the $\gamma$-ray energy.
This paper describes this study comparing data taken during the AGILE calibration occurred
in 2005 with simulation. 

\end{abstract}

\end{frontmatter}

\section{Introduction}

AGILE (Astro-rivelatore Gamma a Immagini LEggero) is a Small Scientific
Mission of the Italian Space Agency (ASI), dedicated to high-energy astrophysics.
It combines two co-aligned imaging detectors operating respectively in the X and
$\gamma$-ray bands with large Field of Views (FoV). 
The Silicon Tracker (ST), at the core of the AGILE satellite, is designed to 
detect and image $\gamma$ rays in the
30 MeV - 50 GeV energy range \cite{sttb}-\cite{st1}-\cite{agimis}-\cite{agimis2}.\\
The on-ground calibration of an astronomical instrument is important for 
the interpretation of its results. 
The goal is to reproduce in laboratory, under controlled condition,
the response of the instrument in flight.
This task requires a tagged photon beam with position, direction and energy of 
each photon known with a precision better than the instrument resolution. 
The realization of such a beam turns out to be a challenging endeavour.
The Beam Test Facility (BTF) in the DA$\Phi$NE collider 
complex at the INFN Laboratori Nazionali of Frascati (LNF) \cite{btfcomm},
was the elected site for realizing the tagged photon beam exploiting
bremsstrahlung in a thin target and performing the calibration.
Preliminary results on the calibration of the AGILE ST has been 
presented in \cite{cattaneo-ricap} and will be subject of a future paper.\\
This paper presents the experimental setup, a detailed Monte Carlo study of the
system and the comparison with the experimental results collected during the calibration. 


\section{The experimental setup}

The experimental setup is a complex system consisting of the BTF e$^-$ beam, the
target to generate bremsstrahlung photons, the spectrometer magnet and 
the detector to measure the energy of post-bremsstrahlung electrons.
The various subsystems are described in the following with the convention that 
$y$ is the coordinate perpendicular to the BTF line plane,
$x$ the one transverse to the beam at the target in the BTF line plane and $z$ is the one
along the beam at the target, that is the direction of the bremsstrahlung photons.

\begin{figure}[htb]
\begin{center}
\includegraphics[angle=0,width=1.0\textwidth]{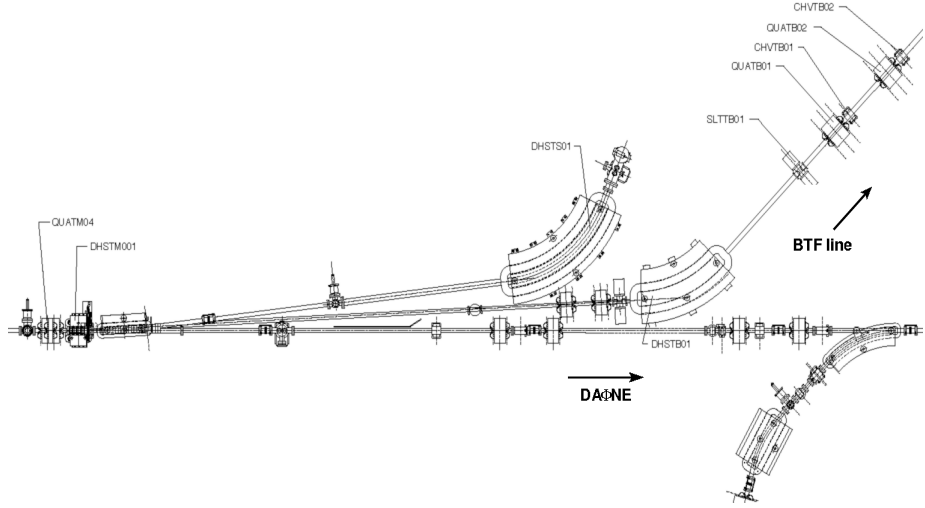}
\end{center}
\caption{The initial section of the BTF transfer line.} 
\label{linea-fascio}
\end{figure}

\subsection{The electron beam}

The e$^-$s used for generating the photon beam are delivered by the BTF. 
The BTF is fed by the DA$\Phi$NE complex that provides 
e$^\pm$ under carefully controlled condition with predefined multiplicity.

\subsubsection{The Beam Test Facility (BTF)}

\begin{figure}[htb]
\begin{center}
\includegraphics[angle=0,width=0.75\textwidth]{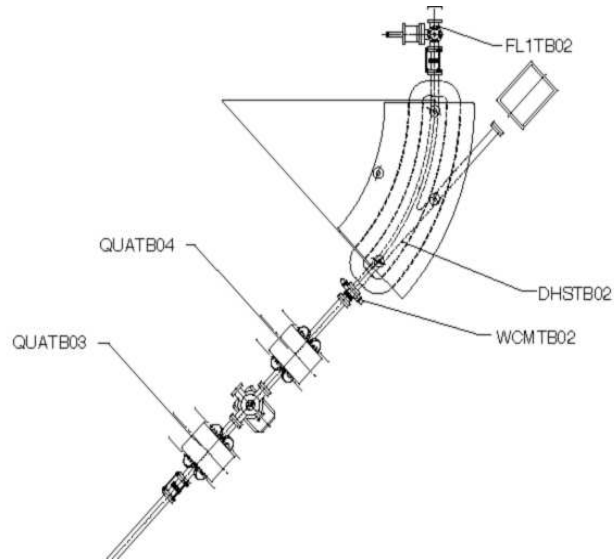}
\end{center}
\caption{The final section of the BTF transfer line including the last spectrometer magnet DHSTB02.}
\label{btf_halla}
\end{figure}

For the ST calibration we used the BTF in the DA$\Phi$NE 
collider complex at LNF, which includes a LINAC at high e$^+$/e$^-$ currents, an accumulator
of e$^+$/e$^-$ and two storage rings at 510 MeV.
The e$^+$/e$^-$ beam from the LINAC is directed into the accumulation ring to be subsequently 
extracted and injected in the Main Ring. When the system injector does not transfer the 
beams to the accumulator, the beam from LINAC can be extracted and transported in the test beam area 
through a dedicated transfer line: the BTF line (Fig.~\ref{linea-fascio}). 
The BTF can provide a collimated beam of e$^+$/e$^-$ in the energy range
20-750 MeV with a pulse rate of 50 Hz. The pulse duration can vary from 1 to
10 ns and the average number of e$^-$ per bunch $\overline N_e$ ranges from 
$1$ to $10^{10}$ \cite{btfcomm}-\cite{btf-ieee-2004}-\cite{btf-epac-2006}.\\
The BTF can be operated in two ways:
\begin{itemize}
\item {\it LINAC mode:} operating when DA$\Phi$NE is off, with a tunable energy 
between 50 MeV and 750 MeV and an efficiency around 0.9
\item {\it DA$\Phi$NE mode:} operating when DA$\Phi$NE is on, with a fixed energy 
at 510 MeV and an efficiency around 0.6
\end{itemize}
The extracted electrons are transported to the BTF hall, where the final section is located (Fig.~\ref{btf_halla});
the experimental equipment under test is positioned at the exit of the spectrometer magnet DHSTB02.
All or part of the equipment can be mounted under vacuum continuing the beam line or, 
alternatively, the beam line can be terminated with a thin window and the equipment mounted in air.\\
In spite of some disadvantage from the point of view of background, this last option was adopted 
with a $0.5\,\mathrm{mm}$ Be window because of the difficulties in implementing an extended vacuum
line incorporating all the required equipment.


\subsubsection{e$^-$ multiplicity per bunch}
\label{mulbun}

The calibration of the ST should be ideally performed in a single-photon 
regime, avoiding simultaneous multi-photon production to reproduce the 
astrophysical conditions.
Multi-photon events should ideally identified and rejected otherwise they
will bias the counting statistics.\\
On the other hand, bremsstrahlung is a continuous process and multi-photon
generation (with photon energy $E_\gamma$ above a given threshold) is possible also when a 
single e$^-$ crosses the target.
The fraction of multi-photon events is approximately proportional to the single-photon
emission probability. That implies the need of a compromise between the photon beam
intensity and the single-photon beam purity.\\
Considering that the target thickness was constrained by the availability 
of the hardware, by the need to guarantee full detection efficiency of the beam
electrons and by the need to measure electrons twice both in the $x$ and $y$ directions
for studying the beam size and divergence, the only free parameter is the 
e$^-$ multiplicity per bunch.\\
Another constraint is the limited time available for the calibration campaign
and the request of calibrating many different ST geometrical configurations.
That puts a lower limit to the required photon flux and therefore on the e$^-$ 
multiplicity per bunch.\\
In DA$\Phi$NE mode with 5 e$^-$/bunch the fraction
of multi-photon events having $E_\gamma > 20\, \mathrm{MeV}$ can be 
estimated to be $\approx 10\%$ by the formulae in the Appendix. 
This uncertainty is greater than the accuracy requirement on the
effective area measurements. On the other hands, the DA$\Phi$NE mode with 1
e$^-$/bunch is consistent with the accuracy requirements but the time necessary
to collect enough statistics is incompatible with
the time available for the calibration. 
The ST cross-section for photons with $E_\gamma < 20\,\mathrm{MeV}$ is 
not negligible and thus the fraction of interacting secondary photons will be 
larger than the numbers calculated in the Appendix.
Taking into account the above considerations, the best configuration for 
ST performance and calibration would be with 1 e$^-$/bunch, but the flux 
requirement forced to select the configuration with $\approx 3$ e$^-$/bunch.

\begin{figure}[htb]
\begin{center}
\includegraphics[width=0.7\textwidth,angle=0]{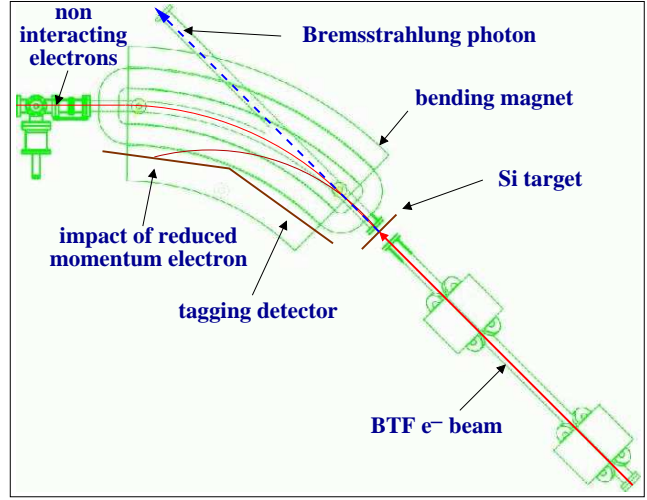}
\end{center}
\caption{A schematic view of the $\gamma$-ray line: the target, the spectrometer magnet and the PTS.}
\label{tagger}
\end{figure}

\subsection{The bremsstrahlung target}
\label{btar}

Photons in the energy range relevant for the ST are produced by
bremsstrahlung of electrons in a target;
subsequently a magnet bends away the electrons while the $\gamma$ rays can 
travel towards the AGILE instrument (see Fig.\ref{tagger}).\\
The bremsstrahlung target consists of two pairs of silicon single sided micro-strip
detectors of area $8.75 \times 8.75\, \mathrm{cm}^2$ and $410\, \mathrm{\mu m}$ thick, 
each including 768 strips with $114\,\mathrm{\mu m}$ pitch. Only every other strip is read, 
so that each target detector has 384 readout channels with $228\, \mu m$ readout pitch. 
Each pair measures separately the $x$ and $y$ coordinates transverse to the beam.
A spatial resolution $\sigma \leq {114}/{\sqrt{12}}\,\mu\mathrm{m} 
\approx 33 \,\mu\mathrm{m}$ is expected; the cluster size is often limited 
to one strip and therefore the resolution is limited by the strip pitch.\\
The target has two roles: to measure the coordinate and the direction
of the electrons and to cause the emission of bremsstrahlung photons.
The target detectors are positioned along the beam direction between the last 
focusing magnet (QATB04 in Fig.\ref{btf_halla}) and the spectrometer magnet
(DHSTB02 in Fig.\ref{btf_halla}). The $x$ measuring ones
are the first and the third, positioned respectively at $5.45\,\mathrm{cm}$ and 
$7.20\,\mathrm{cm}$ downstream the Be window, while the $y$ measuring ones
are the second and the fourth, positioned respectively $6.45\,\mathrm{cm}$ and
$8.20\,\mathrm{cm}$ downstream the Be window.\\
At the electron energy most used during calibration, $E_e=463\,\mathrm{MeV}$, the 
contribution to beam divergence due to Coulomb Multiple Scattering in each target 
detector is evaluated under the Gaussian approximation from \cite{pdb} as 
$\approx 2.0\,\mathrm{mrad}$.

\subsection{The Photon Tagging System (PTS)}

\begin{figure}[htb]
\begin{center}
\includegraphics[width=0.8\textwidth,angle=0]{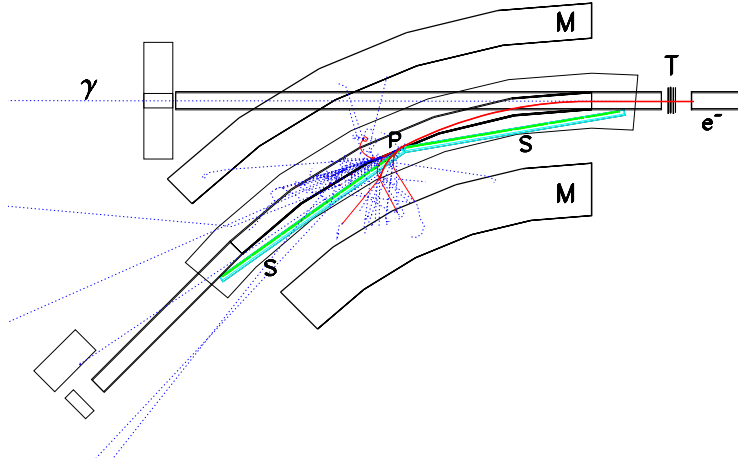}
\end{center}
\caption{Geometry of the spectrometer magnet M drawn in GEANT3 including the 
PTS detectors S displayed with an e$^-$ (entering from the right)
irradiating a $\gamma$ ray in the target T and hitting the PTS at P. 
Photons are represented by dashed (blue) lines and electrons by solid (red) lines.
In this event the $\gamma$ ray energy is $\approx 75$ MeV.}
\label{ptshit}
\end{figure}

The spectrometer magnet, visible in Fig.\ref{btf_halla}-Fig.\ref{tagger}-Fig.\ref{ptshit}, 
generates a dipolar field along the $y$ direction 
over an angular range of 45$^\circ$. In between the two magnet poles, there 
is a pipe made of stainless steel with rectangular section. 
It is composed of a straight section ('photon pipe') along which the 
bremsstrahlung $\gamma$ rays travel to the ST and a curved section 
('electron pipe') defining the trajectory for e$^-$s bent by the 
magnetic field.\\
The pipe is hollow with an air filled inner section with size
$5.50\times 3.50$ cm$^2$, its wall thickness is $0.35\,\mathrm{cm}$. 
The magnetic field in the volume between the poles that includes the 
'electron pipe' is assumed constant with strength 
$B=0.9\,\mathrm{T}$ when $E_e = 463.0\,\mathrm{MeV}$, corresponding 
to a curvature radius $R=172.0\,\mathrm{cm}$.\\
The equipment for the detection of the e$^-$s that lost energy in the target
was developed and installed inside the spectrometer magnet by our team: the 
Photon Tagging System (PTS).
It consists of 12 micro-strip silicon detectors positioned on the internal walls
of the spectrometer magnet (see Fig.\ref{btf_halla}, Fig.\ref{tagger} and Fig.\ref{ptshit}) 
grouped into two modules of 6 detectors each, located into two hollow rectangular 
aluminum boxes few mm thick. In each module, the detectors are 
located along a straight segment and therefore follow only approximately the 
curved section of the 'electron pipe'.
The area of each detector is $11.86 \times 2.00\, \mathrm{cm}^2$ 
with thickness $410\,\mathrm{\mu m}$ and is subdivided in
1187 strips with $100\,\mathrm{\mu m}$ pitch. Only every third strip is read
resulting in 384 readout strips per detector and 4608 in total.\\
Between each pair of consecutive detectors inside a module, there is a gap 
$\approx 6\,\mathrm{mm}$
wide that is effectively a dead area. A larger gap $\approx 2.0\,\mathrm{cm}$ wide 
is present between the two modules that contributes to the dead area as well.
Electronic noise gives a small contribution compared to the signals from e$^-$
amounting to $\approx 2\,\mathrm{keV}$ that is of little relevance for the measurements.
Depending on the energy loss in the target, the electrons impinge on 
different strips. The correlation in time between the signals of the e$^-$ 
in the target and in the PTS tags the photon; the position on the PTS 
measures the photon energy.\\
The trigger for reading out the target and PTS data is given by the delayed LINAC pre-trigger;
it is read out independently from the ST data. This point has great 
relevance for the following analysis.

\section{The Monte Carlo simulation}
\label{geantmc}

A proper characterization of the BTF requires a careful comparison of the data with a
detailed Monte Carlo simulation. The simulation is realized using the GEANT3 package \cite{GEANT3}.
The simulation incorporates a description of the electron beam delivered by the accelerator
complex with beam parameters determined partly from design values and partly from measurements.
The number of e$^-$ per bunch $N_e$ can be fixed to an integer value or can follow a Poisson 
distribution averaged at any real value $\overline N_e$.\\
The material distribution of the bremsstrahlung target and of the spectrometer 
is simulated in detail. The target and the pipe can be simulated
in air or in vacuum in various configurations. The default configuration is the one actually
used during data taking with the target and the pipe in air.\\
The interactions of electrons and photons 
are driven by the GEANT3 routines with the possibilities of switching on and off the relevant
physics processes like Coulomb Multiple Scattering, bremsstrahlung, Compton scattering, pair
creation for all or only for some of the materials. This option turned out to be very useful
in understanding the behaviour of the BTF/PTS system.
The energy cuts for the e$^\pm$ and photons are kept at the minimum allowed by the program 
(100 keV). A gauge of the quality of these cuts is the average energy loss of a minimum 
ionizing particles crossing a $\approx 400\,\mathrm{\mu m}$ thick silicon layer comparable 
to a target detector or a ST layer thickness: it is $\approx 100$ keV. This level of precision
is required to simulate spurious hits in the target and in the ST that can affect the 
measurement.\\
The digitization simulation in the silicon micro-strip detectors is based on a simplified model: 
the charge released in the volume below each strip is collected by the strip without 
accounting for diffusion and charge trapping. Exploiting the capacitive coupling \cite{cc1990}, 
the charges collected on all strips are fed into the readout strips with appropriate 
coefficients as described in \cite{sttb}.
The noise is simulated simply adding a Gaussian distributed charge on each strip 
around a cluster; the width is determined by the data and amount to 
$\approx 2\,\mathrm{keV}$.\\
The e$^-$ beam is generated inside the last
straight section of the accelerator $\approx 5\,\mathrm{cm}$ upstream the target
with a beam spot of elliptical shape with a 2D Gaussian distribution, with angular 
divergences perpendicular to the beam generated according 2 separate Gaussian 
distributions and with a Gaussian distributed momentum spread.
The momentum spread is provided by the DA$\Phi$NE staff, while the other 
beam parameters are directly measured.

\subsection{The simulated photon beam }

\begin{figure}[htb]
\begin{center}
\includegraphics[angle=0,width=0.7\textwidth]{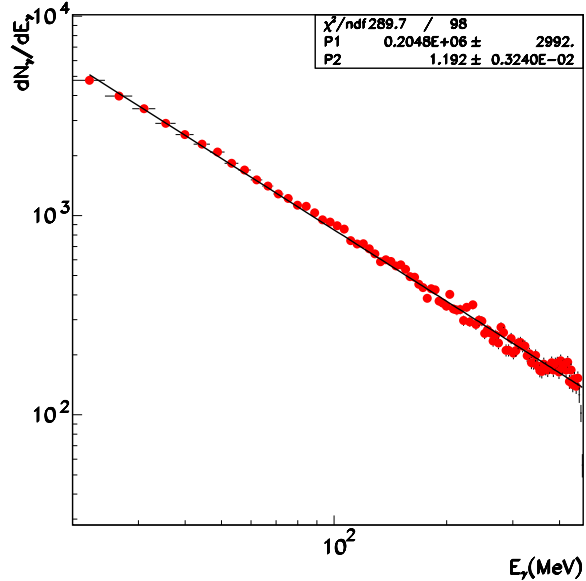}
\end{center}
\caption{Spectrum of the $\gamma$-ray beam reaching the ST fit with a power law $\alpha/E^\beta$.}
\label{sgamma}
\end{figure}

The photon beam directed to the ST is simulated through the interaction of 
the e$^-$ beam with the target. The photon generation is due to the 
bremsstrahlung effect and follows Eq.\ref{sigdy} in the Appendix. This formula 
shows an approximate $1/E_\gamma$ dependency, that is a power spectrum with 
index $\approx -1$. More precisely, a fit of 
Eq.\ref{sigdy} in the interval $0.05-1.00$ with a power spectrum 
returns an index $\approx -1.2$.\\
The simulated energy spectrum of the most energetic $\gamma$ rays 
reaching the ST (but not necessarily interacting with it) is shown in 
Fig.~\ref{sgamma} with a power spectrum fit returning an index $\approx -1.2$ 
as expected by the analytical formula. The number of e$^-$s per bunch 
is set to $N_e=1$ 
without Poisson fluctuation. The same result is obtained when $\overline N_e=3.5$ 
with Poisson fluctuation.
This result is not obvious because of the energy dependent interactions along 
the path that could change the spectrum, that turn out to be small.

\begin{figure}[htb]
\begin{center}
\includegraphics[angle=0,width=0.7\textwidth]{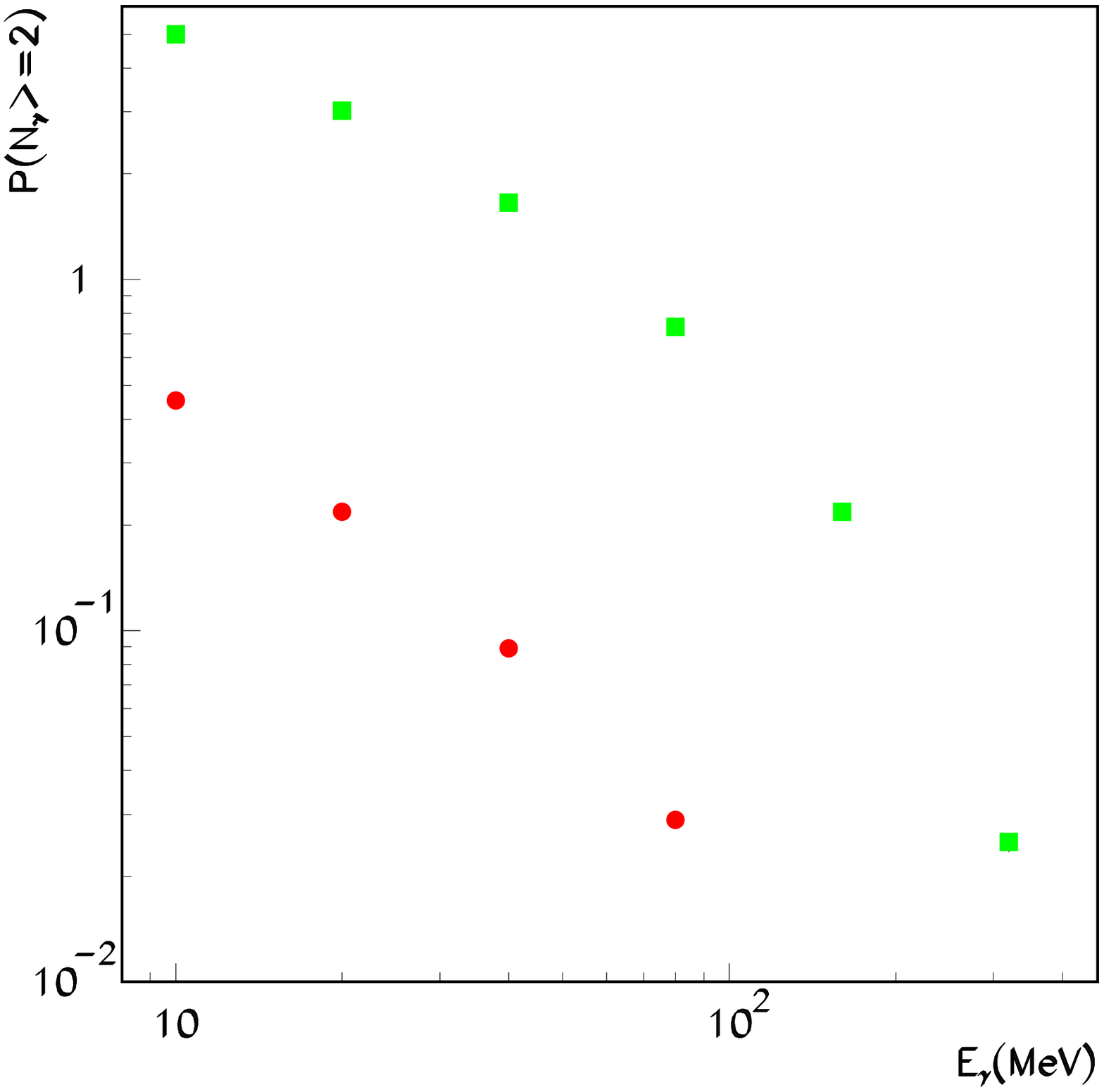}
\end{center}
\caption{Probabilities (in \%) of multi-photon events versus 
$E_\gamma$ threshold for $N_e=1$ (fixed, red dots) and $\overline N_e=3.5$ 
(Poisson distributed, green squares).}
\label{mgamma}
\end{figure}

\noindent
Another important element that characterizes the beam is the fraction of 
multi-photon events. Because in an astrophysical environment there
are no such events, they must be considered background and must be minimized.
If an event has two or more photons with energy above the detection threshold in the 
ST (about few tens of MeV), they can interact simultaneously in the ST. 
The reconstruction software, designed for the astrophysical environment, is not 
fit to identify such events and will most likely fail or return incorrect energy
and direction. The percentage fraction of multi-photon events above a 
threshold is shown in Fig.\ref{mgamma}. Even for events with $\overline N_e=3.5$, the 
fraction is not larger than a few \%\ even for $E_\gamma$ as small as 10 MeV. 
Therefore their contribution to errors in the measurement of the ST performances 
is at most of comparable size.

\noindent
A related but somehow different issue is the number of low energy photons, say
$E_\gamma<10$ MeV, accompanying a photon in the ST energy range ($E_\gamma>30$ MeV). 
This photons are not enough energetic to convert in a e$^+$e$^-$ pair 
detectable in the ST, but they can interact in coincidence with 
a photon of higher energy generating spurious hits that can influence the 
proper reconstruction of the event.

\section{Results and discussion}

The analysis of the target and the PTS data allows the validation of the Monte Carlo simulation
required to correlate the $\gamma$-ray energy and the PTS hit position.

\subsection{The bremsstrahlung target data }
\label{bremdata}

The analysis of the target data starts from looking for strips above a threshold, 
then neighbouring strips are grouped into clusters. The cluster coordinate is 
obtained calculating its centroid by weighting each strip coordinate with 
the charge collected.\\
Ideally, each cluster should signal a hit of one e$^-$ in a target detector
and the passage of each e$^-$ should be signaled by one hit in each 
target detector. In practice there is the possibility of inefficiency in the detection of 
e$^-$ hits, noise hits, multiple clusters due to a single e$^-$ and single clusters 
on the same target detector associated to multiple e$^-$s.\\
The efficiency of the hit searching algorithm is measured by selecting the events 
with one hit on three target detectors and zero on the fourth. This number is to be compared 
with the events with one hit on each target detector.
With the available statistic, no event satisfies this requirement, so that 
the efficiency is basically $100\%$.\\
The fraction of noise induced hits can be estimated by requiring one hit on one target detector
and zero hit on the others. Also in this case, no event satisfies this requirement and 
therefore the fraction of noisy hits is basically $0\%$.\\
Particularly interesting samples are the 0-cluster events, where no cluster is detected in 
the target, and the 1-cluster ones, where one cluster per target detector is detected. For the 
considerations detailed above, the first sample can be safely associated to events with 
0-e$^-$ events; even these events are triggered because the
target signal is not required by the trigger logic.
The 1-cluster sample mostly overlaps with 1-e$^-$ events and is used in the following
for characterizing the beam.
The main goal of these measurements is inferring the beam 
properties to be used in the Monte Carlo generation. 

\begin{figure}[htb]
\begin{center}
\includegraphics[angle=0,width=1.\textwidth]{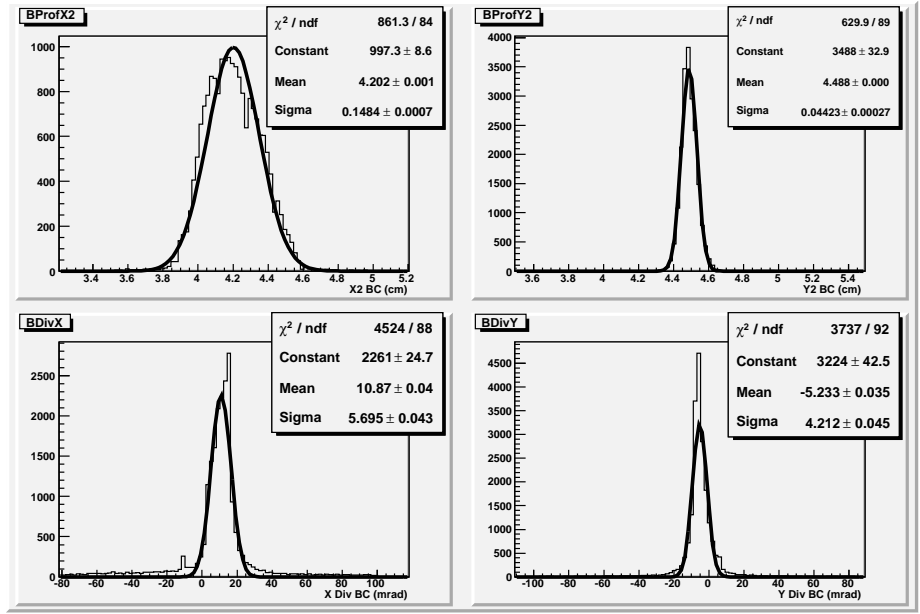}
\end{center}
\caption{Beam parameters at $E_e=463$ MeV measured on data: 
$x$ profile (top left), $y$ profile (top right), $x$ divergence (bottom left),
$y$ divergence (bottom right).}
\label{beamprofile}
\end{figure}

\subsubsection{Beam sizes}

The beam sizes are measured only with 1-cluster sample. 
The beam profiles using the first $x$ and $y$ measured coordinates in the 
target are shown in the top row of Fig.\ref{beamprofile} with the results
of Gaussian fits superimposed.
Using the second $x$ and $y$ 
measured coordinates in the target gives compatible beam sizes.\\
The beam sizes are $\sigma_x \approx 1.5\,\mathrm{mm}$ and 
$\sigma_y \approx 0.5\,\mathrm{mm}$ with a significant non Gaussian tail in $y$.
This numbers are representative but subject to significant variations for
different runs, due to changes in the beam setting.

\subsubsection{Beam divergences}

Angular divergences are measured using the 1-cluster sample. The estimator is 
the difference between the cluster coordinates ($x_i(y_i),i\in (1,2)$) on the 
target divided by their distance $d_{x(y)}$
\begin{eqnarray}
\label{sxy}
s_x = (x_2 - x_1)/d_x \nonumber \\
s_y = (y_2 - y_1)/d_y 
\end{eqnarray}

\begin{figure}[htb]
\begin{center}
\includegraphics[width=0.95\textwidth]{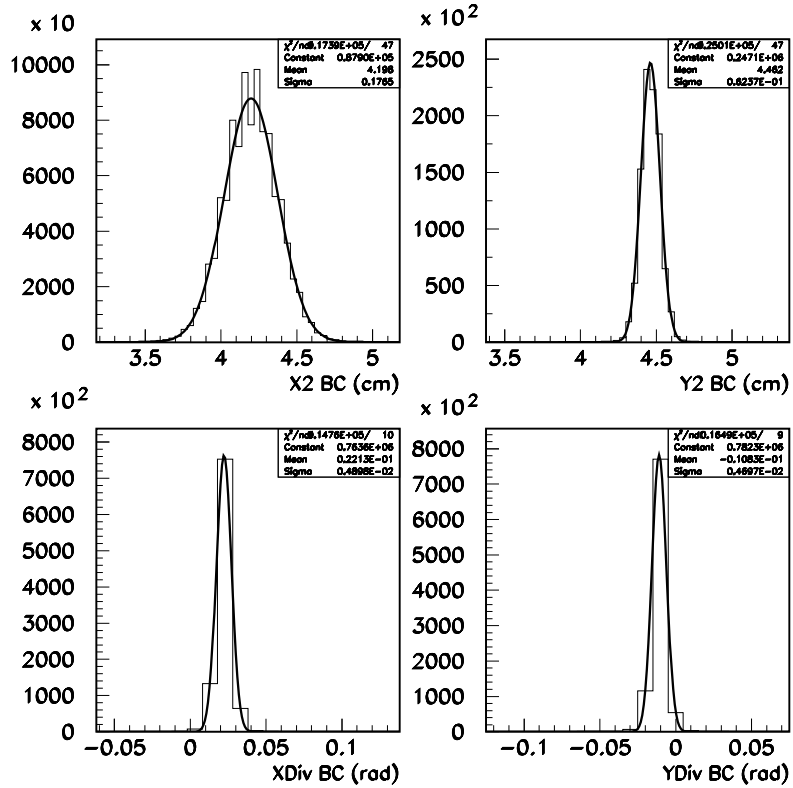}
\caption{Beam parameters at $E_e=463$ MeV measured on a Monte Carlo sample:
$x$ profile (top left), $y$ profile (top right), $x$ divergence (bottom left),
$y$ divergence (bottom right).}
\label{beamprofilemc}
\end{center}
\end{figure}

\noindent 
In Fig.\ref{beamprofile} (bottom row) the distributions of $s_x$ and $s_y$ are shown. 
The Gaussian fit returns $\sigma(s_x) \approx 5.7\,\mathrm{mrad}$ and
$\sigma(s_y) \approx 4.2\,\mathrm{mrad}$.\\
The beam divergences are, especially in $y$, smaller than those presented in 
\cite{btf-epac-2006}. That is due to an optimized setting of the accelerator slits
for this particular application.\\
The measured values for both the sizes and the divergences of the beam are fed into 
the Monte Carlo generator to reproduce the experimental
distributions as shown in Fig.\ref{beamprofilemc}.

\subsubsection{e$^-$ multiplicity }

\begin{figure}[htb]
\begin{center}
\mbox{\begin{tabular}[t]{cc}
\subfigure[]{\includegraphics[width=0.5\textwidth]{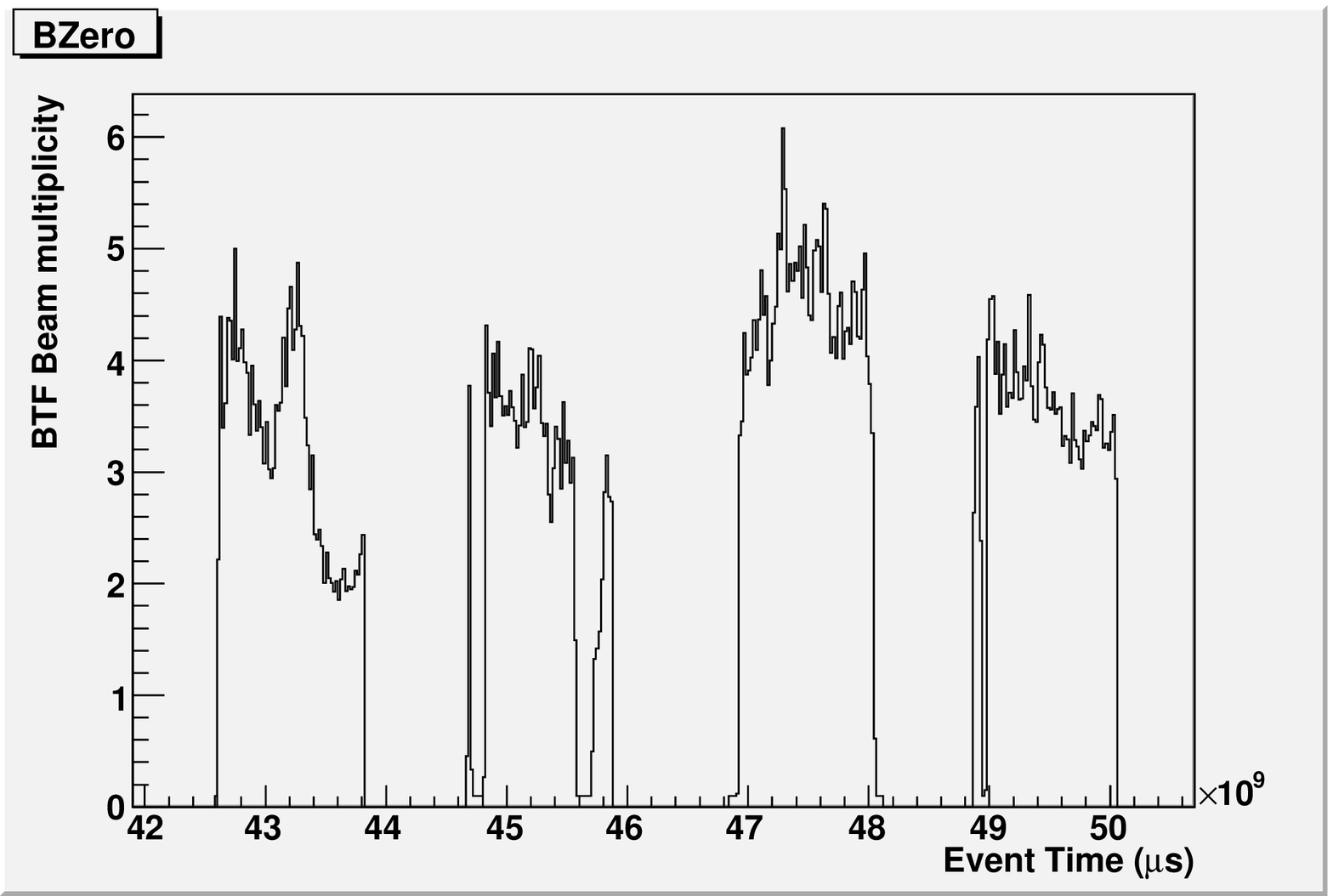}
} &
\subfigure[]{\includegraphics[width=0.5\textwidth]{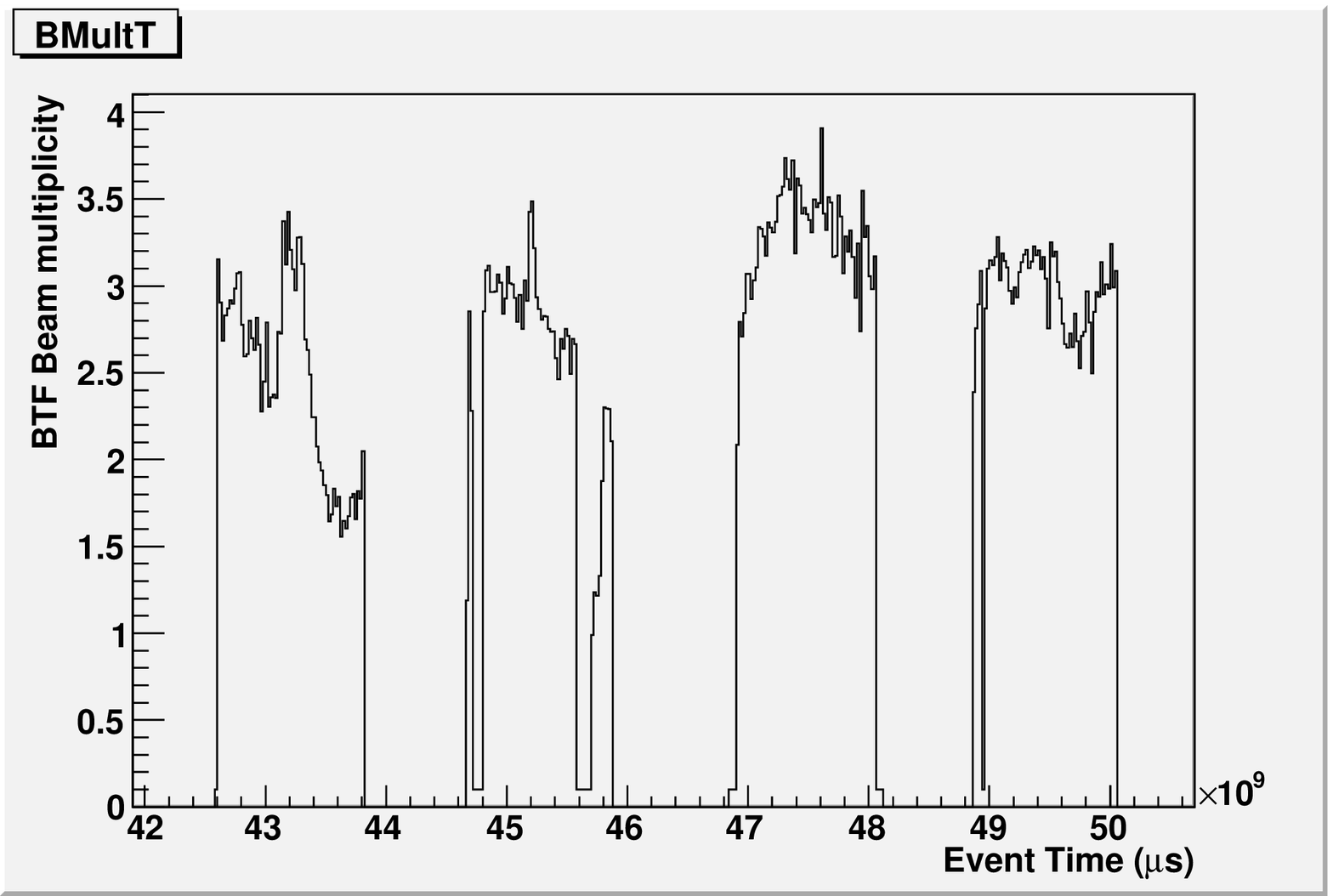}
} \\
\subfigure[]{\includegraphics[width=0.5\textwidth]{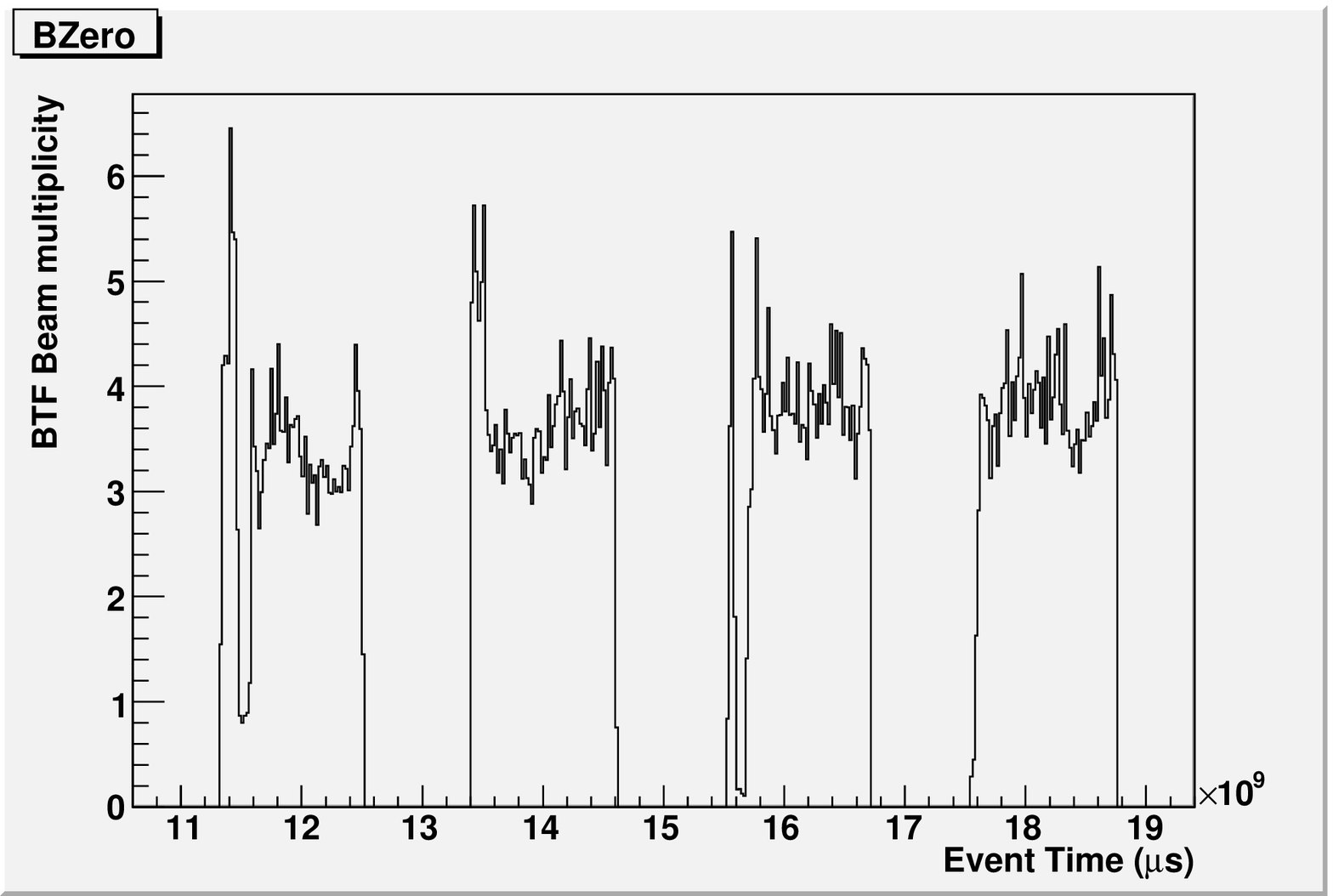}
} &
\subfigure[]{\includegraphics[width=0.5\textwidth]{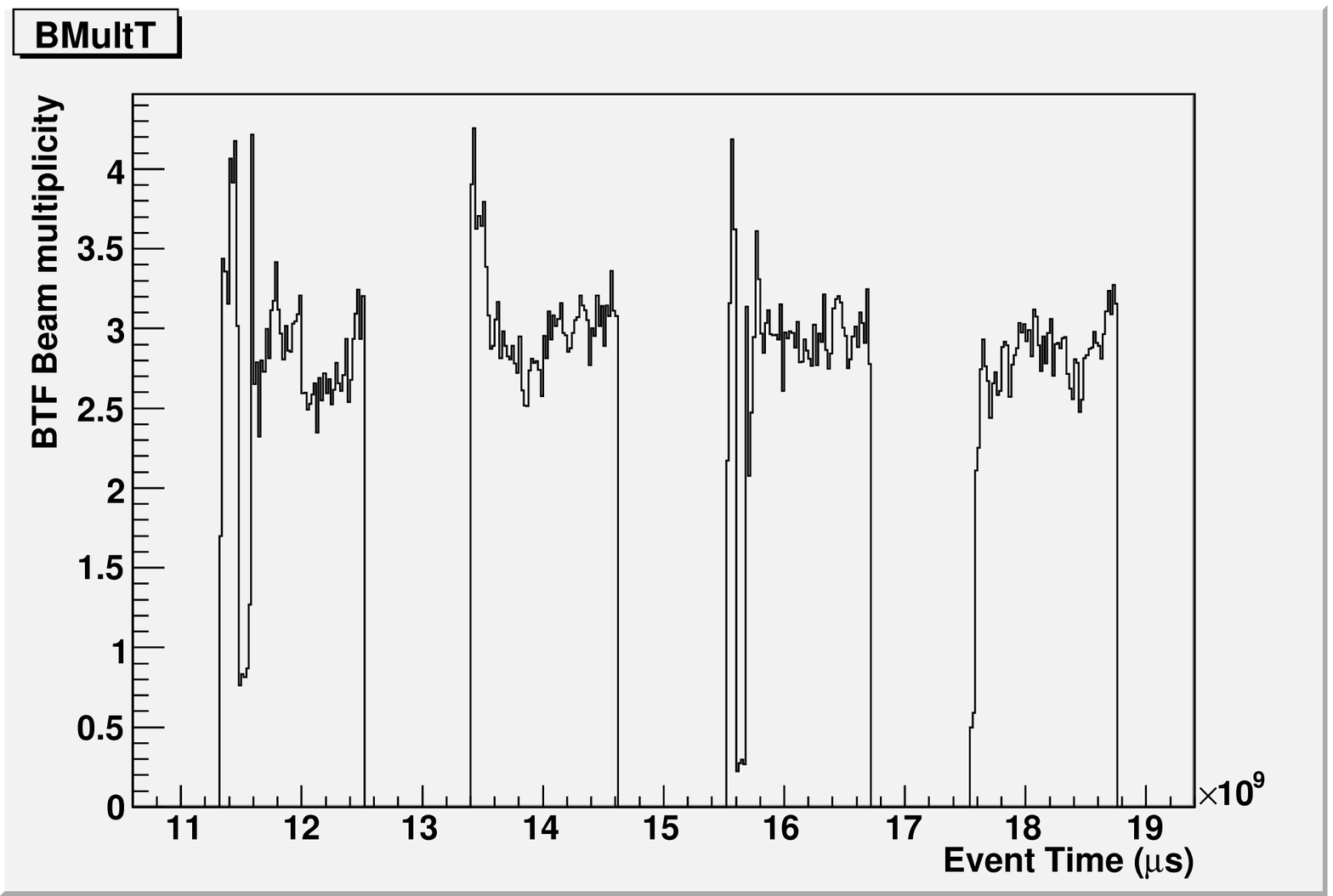}
} \\
\end{tabular}}
\caption{Estimation of e$^-$ multiplicity with the 0-e$^-$ (left) and 
1-e$^-$ (right) samples in the runs 
2328 (top), 2566 (bottom).}
\label{bzero}
\end{center}
\end{figure}

The e$^-$ multiplicity of the BTF beam is one of the most important parameters required for 
an appropriate simulation of the photon beam. As discussed in Sect.\ref{mulbun} 
the average of the Poisson distributed e$^-$ multiplicity is tuned by the accelerator staff.
It is nevertheless important to monitor the e$^-$ multiplicity versus time to assess the reliability
of the assumed multiplicity and its stability in time.
The monitoring is possible by analyzing the hit multiplicity on the target.
The e$^-$ multiplicity is expected to follow a Poisson distribution 
with average $\mu$ 
\[
P_\mu(n) = \frac{\mu^n}{n!} e^{-\mu}
\]
where $\mu$ can be estimated from the fraction of 0-e$^-$ events given by $P_\mu(0) = e^{-\mu}$, that is
$\mu = -\log(P_\mu(0))$.\\
Another estimation is obtained from the ratio of 1-e$^-$ events with 
multi-e$^-$ events 
\[
f_{12} = \frac{\mu e^{-\mu}}{\Sigma_{n>=2} \frac{\mu^n}{n!} e^{-\mu}}
\]
If $f_{12}$ is measured, this relation can be inverted numerically to obtain $\mu$.\\
As previously mentioned 
the 0-e$^-$ events are readily identified as events with 0-cluster events, 
while the 1-e$^-$ sample largely overlaps with 1-cluster events.\\
In order to study e$^-$ multiplicity variation, events can be grouped in subsets 
of 1000 events with a sufficiently 
small statistical error. The estimated 
e$^-$ multiplicity is plotted versus the event time
for two calibration runs in Fig.\ref{bzero}.

These results show a broad agreement with the expected e$^-$ multiplicity but also a quasi 
periodic duty cycle of beam on-off. When beam is on, there are occasional spikes and 
sometimes smoother variations. The estimations from $f_{12}$ is somehow lower than that from 
0-e$^-$ events. That is expected if the chance that a multi-e$^-$ event appears as a
1-cluster event is higher that the reverse. That is what is expected for a narrow beam with 
some chance of overlapping hits from different e$^-$s.

\subsubsection{Beam simulation and comparison with data}

The data contained from the target are used to tune the Monte Carlo 
simulation of the beam. The relevant parameters are the coordinates of the 
beam center, the widths of the beam spot, the beam divergences, the e$^-$ 
multiplicity and the noise.\\
The measurements are performed on the target while the parameters of the beam
in the Monte Carlo generator are generated upstream (see Sect. \ref{bremdata}). 
These parameters may differ because
of the spread of the uncollimated beam and of the multiple scattering along the path.\\
Nevertheless the simulation guarantees that the beam widths at the origin of the Monte 
Carlo generator show no significant difference with those measured on the target. 
Therefore the measured widths $\sigma_x$ and 
$\sigma_y$ can be directly used for the generation of the beam spot.\\
The beam divergences measured with the target are comparable with the 
angular multiple scattering contribution from each of them (see Sect.\ref{btar}). 
Therefore this contribution must be subtracted from the measured one to obtain the values
to be used in generation $\sigma(s_x) \approx 3.6\,\mathrm{mrad}$ and
$\sigma(s_y) \approx 2.7\,\mathrm{mrad}$ with $s_{x,y}$ defined in Eq.\ref{sxy}.
The momentum spread is provided by the DA$\Phi$NE staff as $\sigma(p) = 
5.0\,\mathrm{MeV/c}$.

\subsection{The PTS data }

The PTS data are those tagging the production of a $\gamma$ ray directed to the ST and
providing an estimation of its energy. Ideally the emission of a bremsstrahlung 
$\gamma$ ray from the target would result in an electron hitting the PTS
delivering one cluster of neighbouring strips above a predefined threshold.
In this case the $\gamma$-ray energy would be univocally correlated with the PTS
hit position.\\
The relation between $\gamma$-ray energy and PTS hit position, estimated by the cluster
centroid, is highly non-linear and must 
be calibrated. The calibration can be performed either with the help of analytical 
calculation or relying on Monte Carlo simulations. The former approach cannot easily 
account for the other interactions of the electron in addition to the bremsstrahlung 
in the target. Therefore the $\gamma$-ray energy, $E_\gamma$, is estimated with 
the PTS energy estimator, $E_{PTS}$, calibrated with the
Monte Carlo simulations detailed in Sect.\ref{ptsmc}.\\
Another important point is the treatment of events with multiple PTS clusters. 
Ideally these are multi-photon events and should be discarded from 
the calibration sample. Yet, most 
of the multi-cluster events are genuine single-photon events where additional
clusters are generated by secondary interactions. Furthermore for $E_\gamma \approx
100$ MeV, the e$^-$ trajectories intercept PTS detectors on both modules, resulting
in high probabilities of multi-cluster events even for single-photon events.
Therefore also multi-cluster events are retained.

%

\begin{figure}[htb]
\begin{center}
\includegraphics[angle=0,width=0.8\textwidth]{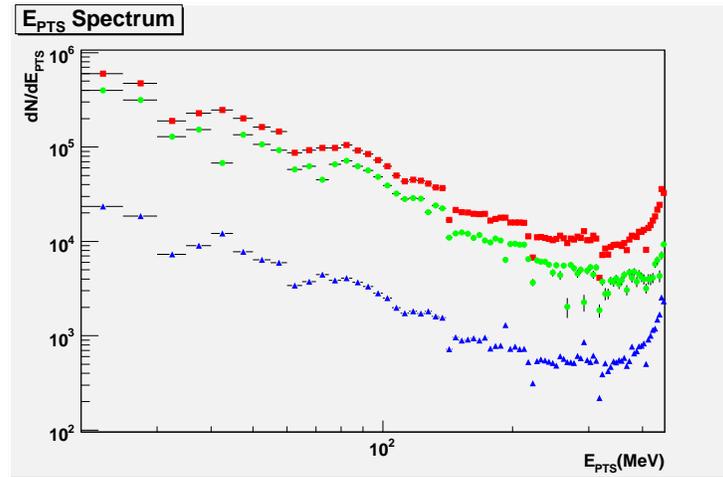}
\end{center}
\caption{PTS energy spectrum: all events (red squares), only 0-e$^-$ events (blue triangles), 
normalized difference (green dots).}
\label{ptsvsegnt}
\end{figure}

\noindent
The spectrum of $E_{PTS}$ is shown in Fig.\ref{ptsvsegnt}.
It displays an anomalous feature at high energy, close to the kinematic limit,
where the increase with the energy looks incompatible with the expected behaviour
of the bremsstrahlung spectrum in Eq.\ref{sigdy}.\\
Two possible explanations are compatible with this excess background: a production
of non-bremsstrahlung related low energy electrons entering into the spectrometer 
or a photon background in phase
with the BTF cycle (the rate of cosmic ray induced hits being too low).\\
The source of the low energy electrons could be $\delta$ rays produced in the interaction
of the beam electrons with the target. They should be proportional to the 
number of electrons in the bunch and follow an approximate $1/E^2_\delta$ spectrum, 
where $E_\delta$ 
is the energy of the $\delta$ electron. This spectrum would fake a $1/(E_e - E_{\gamma})^2$
spectrum
that would qualitatively matches the high energy tail of Fig.\ref{ptsvsegnt}.
On the other hand, Monte Carlo simulations and analytical calculations
show that this source is largely insufficient to account for the excess of PTS hits 
corresponding to high $E_{PTS}$.\\
Furthermore the two sources can be discriminated by looking at 0-e$^-$ events, where no 
PTS hits from beam related $\delta$ rays are expected. Fig.\ref{ptsvsegnt} shows a
significant fraction of 0-e$^-$ events with PTS hits strongly peaked at 
high $E_{PTS}$, that means at low strip numbers.\\
Hence only the latter background source is compatible with all experimental data. 
The existence of a low energy photon background was also confirmed by the accelerator staff.\\
The spectrum of bremsstrahlung $\gamma$ rays can be recovered by subtracting the 
background spectrum appropriately rescaled with the 0-e$^-$ rate but the background events 
cannot be tagged and removed on a event by event basis.

\subsubsection{PTS simulation and comparison with data}
\label{ptsmc}

\begin{figure}[htb]
\begin{center}
\mbox{\begin{tabular}[t]{cc}
\subfigure[]{\includegraphics[width=0.5\textwidth]{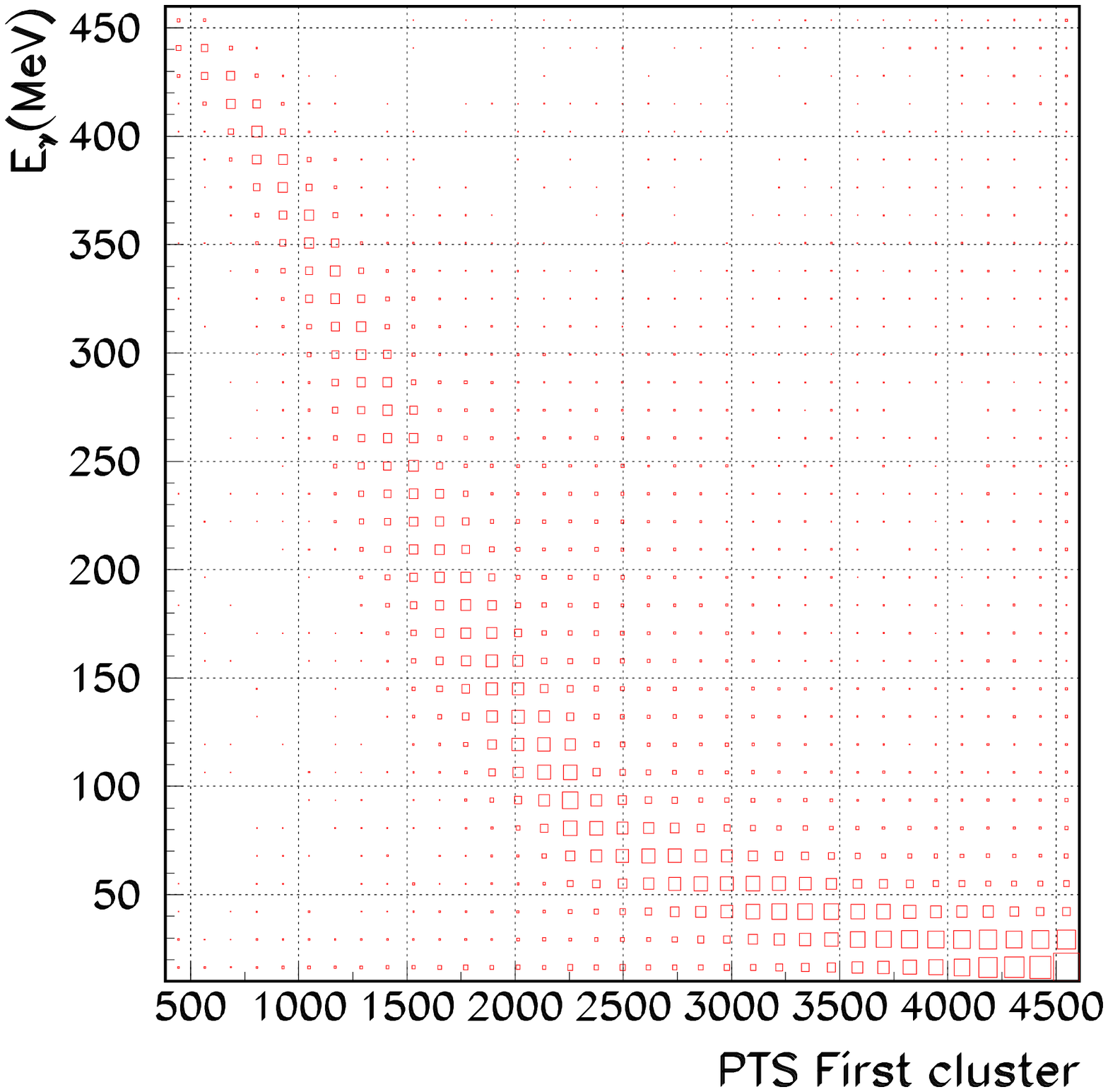}
\label{pts2d}
} &
\subfigure[]{\includegraphics[width=0.5\textwidth]{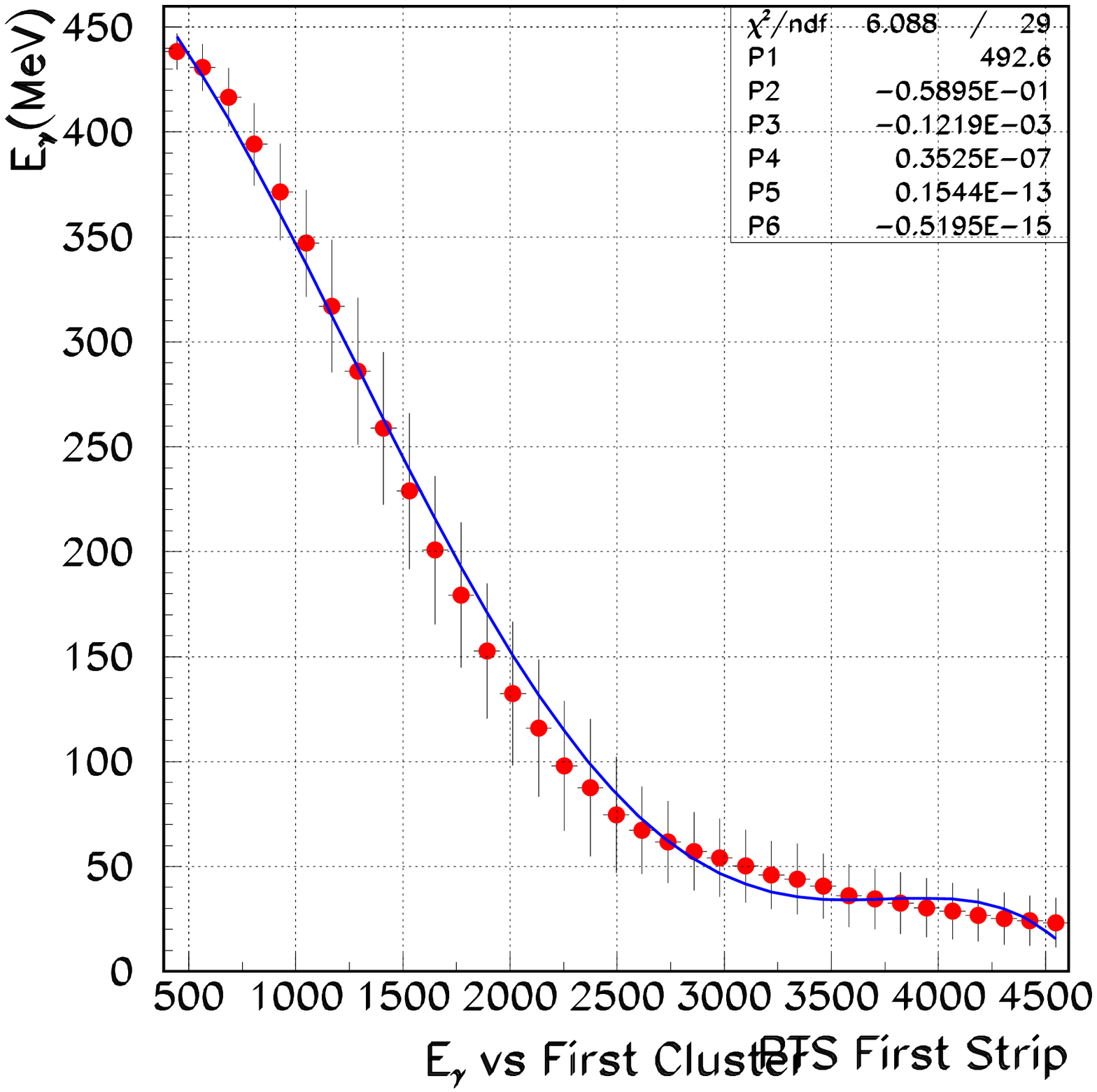}
\label{ptsfit}
} \\
\end{tabular}}
\end{center}
\caption{$E_\gamma$ versus PTS strip number: a) two dimensional b) fitted with 5-th order polynomial.}
\end{figure}

\begin{figure}[htb]
\begin{center}
\includegraphics[angle=0,width=0.75\textwidth]{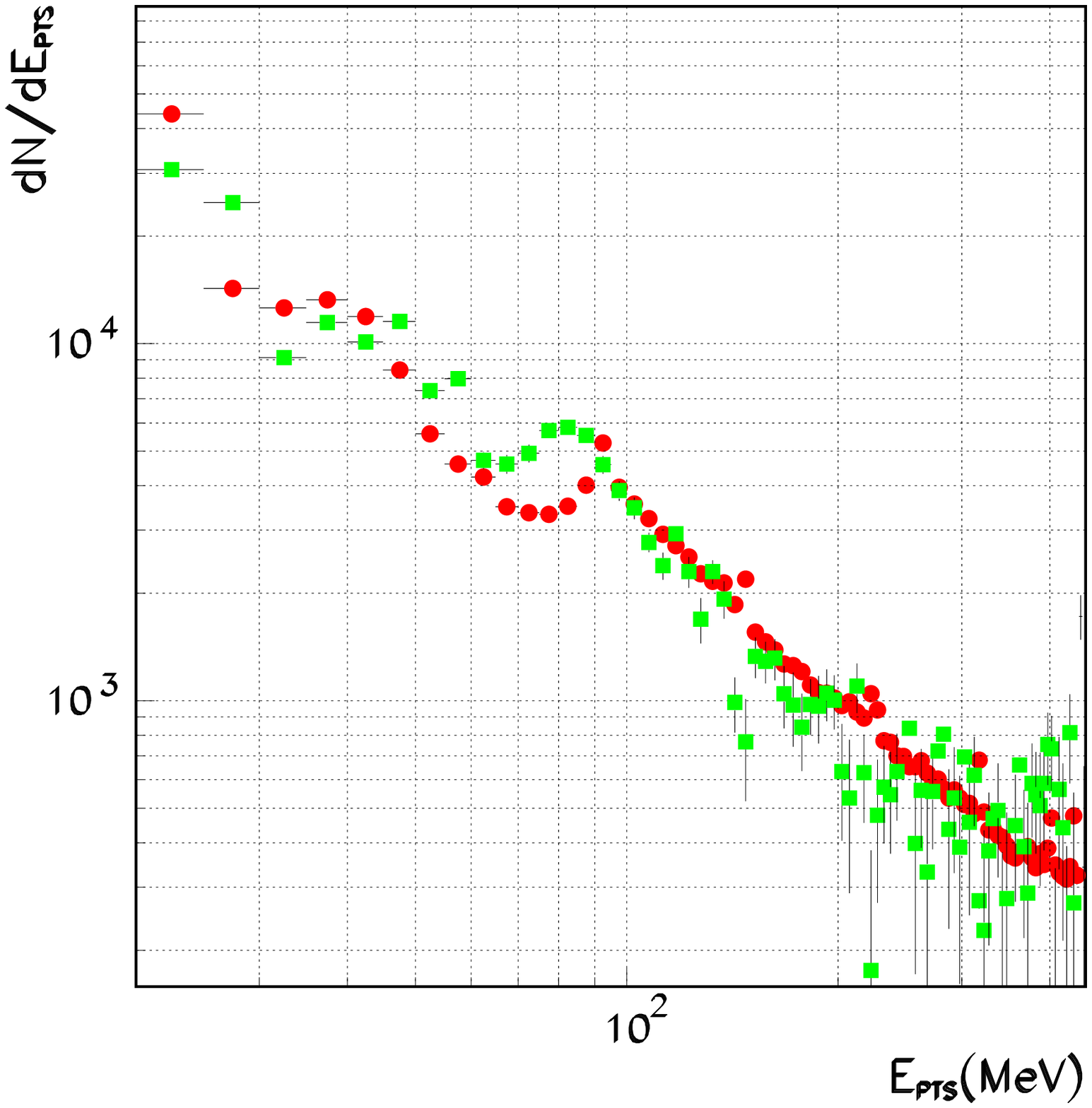}
\end{center}
\caption{PTS energy spectra for data (green square) and Monte Carlo (red circle)}
\label{ptsdatamc}
\end{figure}

In order to provide an interpretation of the PTS data and use them to characterize
the PTS/BTF system, a detailed Monte Carlo study is required. The Monte Carlo data 
contains together with the PTS measurements also the true value at the generation 
level of the particles (Monte Carlo truth), that can be compared to extract the PTS performances.\\
A crucial element is the calibration curve relating the PTS cluster position 
and the electron ($\gamma$-ray)
energy already used in previous sections. The calibration is obtained 
looking at Monte Carlo events generated with 1e$^-$ per bunch by plotting 
the first strip (lowest number) of the first PTS cluster versus $E_\gamma$, 
the energy of the most energetic $\gamma$ ray as shown in Fig.\ref{pts2d}. This plot is 
obtained requiring one PTS cluster, $E_\gamma> 10$ MeV and additional energy from 
$\gamma$ rays other than the most energetic one $<10$ MeV.\\ 
The points are distributed along a band plus a small set of outliers, whose origin 
will be discussed later; in order to use only the points in the band, an additional
graphical cut is applied before the 2D histograms is plotted in form of profile histogram. 
The profile histogram is fitted with a 5-th order polynomial as shown in Fig.\ref{ptsfit}.
The fit has been also tested on simulated events produced with $\overline N_e = 3.5$ 
with Poisson distribution, by allowing multiple PTS clusters or by requiring only
one $\gamma$ ray above 100 keV; the fit results are insensitive to the cuts.
This functional dependence defines $E_{PTS}$ as the PTS measured energy.
The comparison between the data and Monte Carlo $E_{PTS}$ distributions is shown in Fig.\ref{ptsdatamc}.
The agreement is reasonable even if some features appear systematically shifted.

\subsubsection{PTS inefficiency, false positive and Outliers}

The fit in Fig.\ref{ptsfit} is obtained using only events where the PTS energy measurement 
and the true $\gamma$-ray energy are significantly correlated. There are events in which 
that is not the case, that can be divided in three broad classes: PTS inefficiency,
False positive and Outliers.

\paragraph{Inefficiency }

\begin{figure}[htb]
\begin{center}
\includegraphics[width=0.9\textwidth]{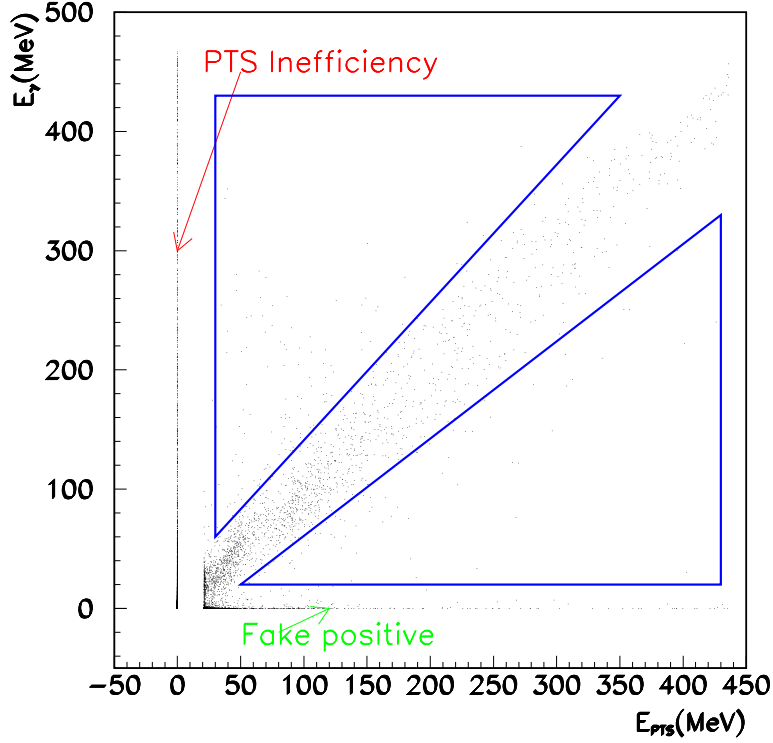}
\end{center}
\caption{$E_\gamma$ vs. $E_{PTS}$ scatter plot. The events far from the central 
band are divided into classes. See text for more details.}
\label{egpts2d}
\end{figure}

The plot $E_\gamma$ vs. $E_{PTS}$ for Monte Carlo events with 1-e$^-$ per bunch is shown 
in Fig.\ref{egpts2d}.
The PTS inefficiencies, displayed in the vertical line on the left, are events where an electron 
emits a $\gamma$ ray by bremsstrahlung but it does not reach the PTS detectors. Such inefficiencies are understood
considering the path of an electron inside the 'electron pipe': in absence of a focusing magnet
and subject to multiple scattering due to the target (and to the air along the path) the electrons 
diverge from the ideal trajectory and may hit the wall of the pipe. That is particularly 
easy along the $y$ direction where the pipe inner half-height is only 
1.75 cm. In this case the e$^-$ showers and the shower particles (e$^\pm$s and photons) may 
hit or not the PTS detectors.\\
Another source of inefficiency originates when the e$^-$ hits the $3.5\,$ mm thick steel inner face
of the guide, as expected, and develops a shower fully in the iron. 
A third cause is due to the e$^-$ hitting the inner face as expected but the 
particles exiting from the outer face do not hit the PTS detectors that have an half-height
of only 1.00 cm.

\paragraph{False positive}

The main cause of false positives, that is $E_{PTS}>0$ and $E_\gamma \approx 0$, is the first 
reason detailed before to explain inefficiencies: an e$^-$ that has not emitted a bremsstrahlung 
$\gamma$ ray showering in the electron
pipe on the top (bottom) face with some particles hitting the PTS detectors. 
The probability of such events is proportional to the extent of the e$^-$ divergence and therefore 
to the e$^-$ path length. Therefore it should increase almost linearly with the strip number, that is
inversely with $E_{PTS}$.\\
False positive may also be the results of $\gamma$ rays actually produced by bremsstrahlung and absorbed
along their path in air or by e$^-$ emitting bremsstrahlung $\gamma$ rays in air along its path
in the electron pipe.

\paragraph{Outliers}

Outliers are generated by a combination of processes similar to those described above.
An e$^-$ emitting a $\gamma$ ray and then hitting the top (bottom) face of the pipe and delivering
a PTS signal generates an outlier.
Outliers can be generated also when the bremsstrahlung 
$\gamma$ ray interacts with air creating a e$^+$e$^-$ pair, that in turn irradiates a lower energy
$\gamma$ ray such that $E_\gamma<E_{PTS}$.\\
In presence of multiple e$^-$s per bunch, a combination of a false positive and of an 
inefficiency generates a outlier.\\
Another relevant source is e$^-$s that, after having emitted a $\gamma$ ray, cross the inner face
of the pipe showering without delivering a PTS signal close to the crossing point. 
The shower photons may nevertheless convert in the PTS detectors at higher strip number,
so that  $E_{PTS}<E_\gamma$.

\subsection{PTS simulation results}

Starting from the complex picture previously discussed the PTS can be 
characterized in different ways. The first step is understanding which 
is the best requirement on the number of PTS clusters. The possibilities are: 
1 cluster, 2 clusters and $\ge 1$ clusters.
The first is expected to have low efficiency but also small number of 
outliers and better energy resolution, the latter has the opposite features 
while the second is a compromise.

\begin{figure}[htb]
\begin{center}
\includegraphics[width=0.75\textwidth]{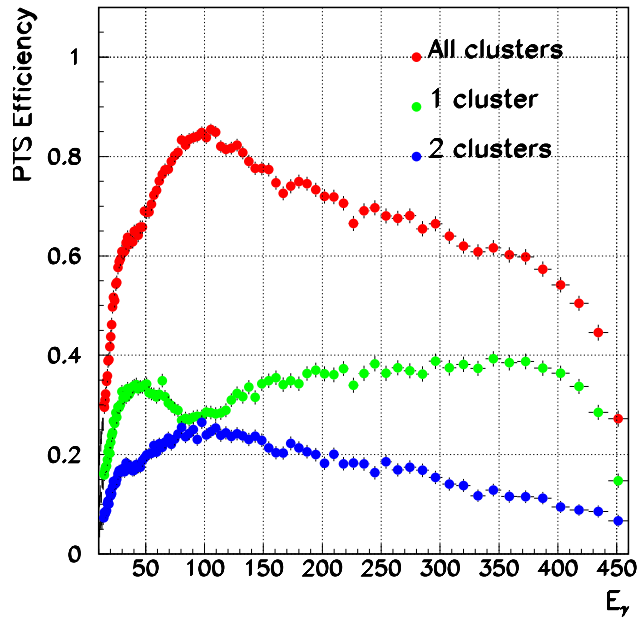}
\end{center}
\caption{PTS efficiency for events with 3e$^-$ per bunch following Poisson 
distribution versus $E_\gamma$.}
\label{ptseff}
\end{figure}

\begin{figure}[htb]
\begin{center}
\mbox{\begin{tabular}[t]{cc}
\subfigure[]{\includegraphics[width=0.50\textwidth]{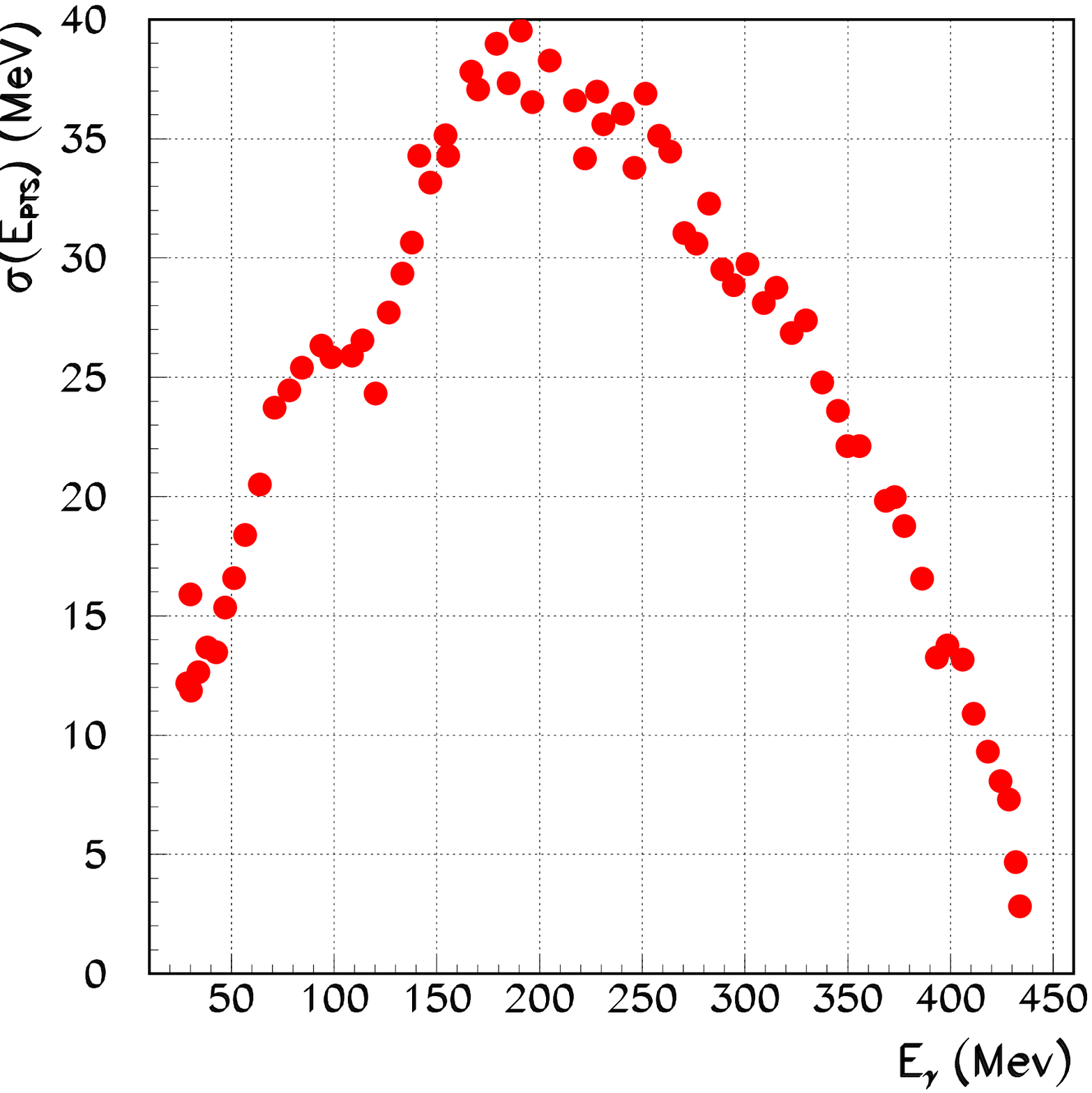}
\label{sigpts}
} &
\subfigure[]{\includegraphics[width=0.50\textwidth]{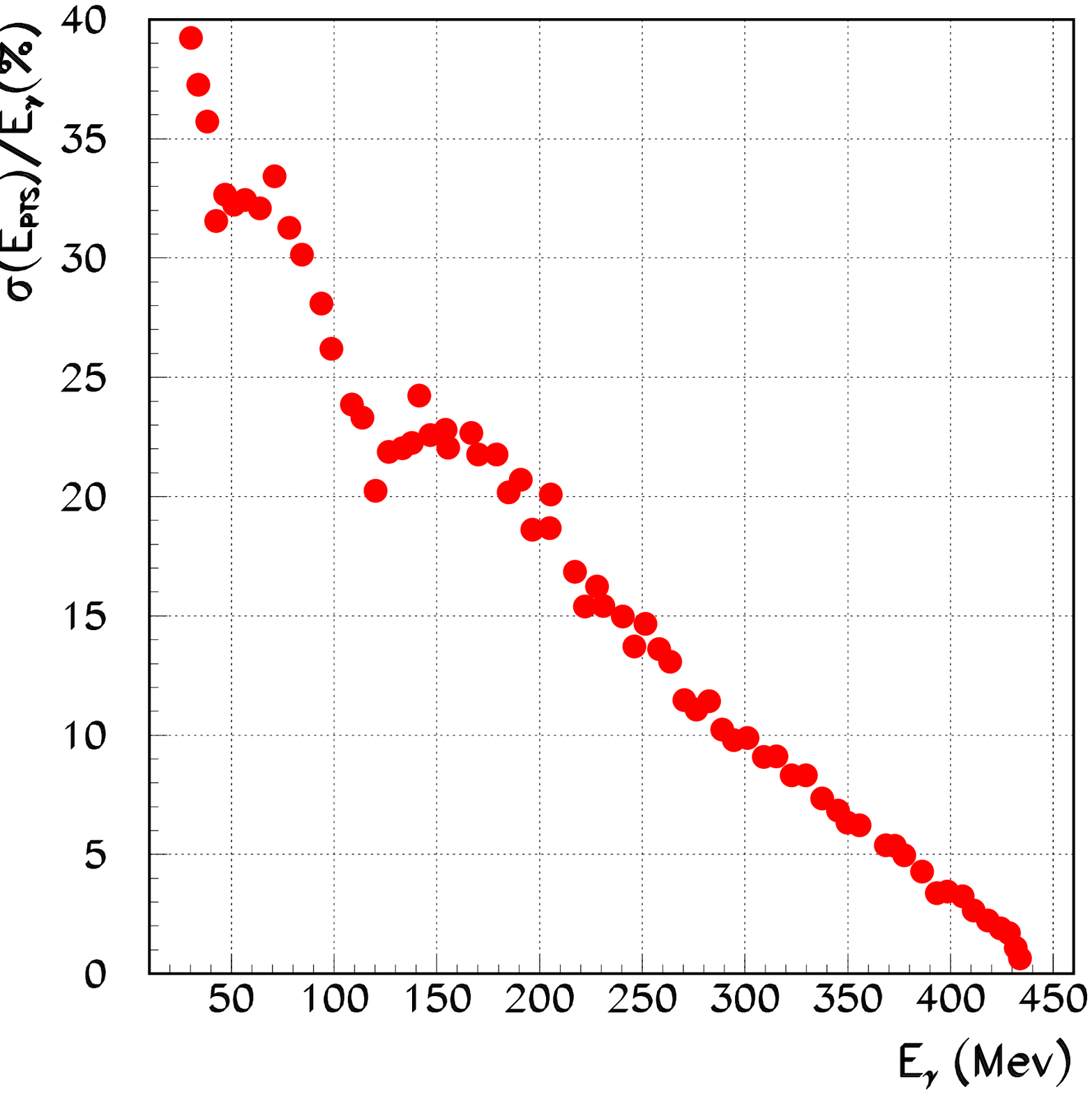}
\label{sigrelpts}
} \\
\end{tabular}}
\end{center}
\caption{Absolute a) and relative b) spread in $E_{PTS}$ distribution for events 
with 3e$^-$ per bunch following Poisson distribution versus $E_{\gamma}$.}
\end{figure}

\noindent
In Fig.\ref{ptseff} the PTS efficiencies for the three cases are shown. 
The option with $\ge 1$ clusters seems to be preferred not only because 
it has higher efficiency but also because it is less sensitive to the 
presence of secondary clusters generated by showering particles.
In this plot a loose definition of efficiency is used that requires only a 
PTS cluster regardless of the $E_{PTS}-E_{\gamma}$ relation, that is 
also outliers are included.\\
The absolute and relative RMS spreads in the $E_{PTS}$ versus $E_\gamma$ 
distribution limited to the central band, that is excluding outliers,
are shown in Fig.\ref{sigpts}-\ref{sigrelpts}. 

\noindent
Another approach is looking at the fraction of events with 
a PTS cluster associated to an energetic $\gamma$ ray versus $E_{PTS}$. That
measures the probability of fake positive with the PTS as a photon tagger. A problem with this 
definition is that there is no infrared limit to $E_\gamma$. The lower limit 
is set by the GEANT threshold for $\gamma$ production. A more 
robust definition is to set a threshold defined by the lowest $E_\gamma$ for which
the PTS has a reasonable efficiency, that is ${\cal O}(10\mathrm{MeV})$. 

\begin{figure}[htb]
\begin{center}
\includegraphics[width=0.75\textwidth]{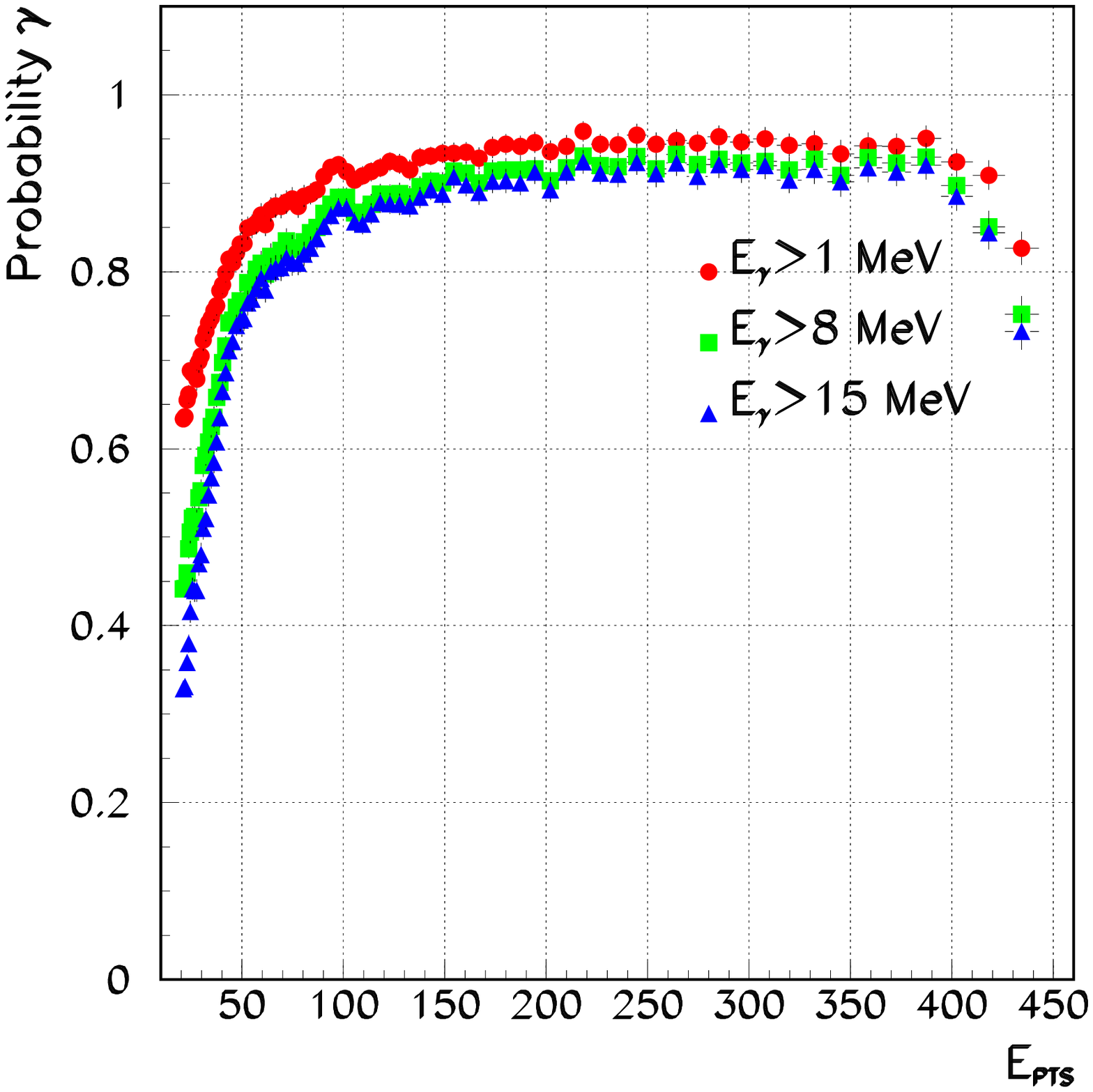}
\end{center}
\caption{Probability of having $E_\gamma>1 (8)(15) MeV$ for events with 3e$^-$ 
per bunch following Poisson distribution versus $E_{PTS}$.}
\label{gameff}
\end{figure}

\noindent
In Fig.\ref{gameff} the probability of having a $\gamma$ ray above the given threshold 
versus $E_{PTS}$ is reported for events with $\ge 1$ PTS cluster. The complement
of this plot gives the fraction of false positive. This fraction is understandably 
high for low $E_{PTS}$ where the probability of a non-emitting e$^-$ hitting the PTS
is high and for high $E_{PTS}$ where the false positives originate from background.
For most of the energy range the false positive 
fraction is $\le 10$\%\ and weakly dependent on the $E_\gamma$ threshold.

\begin{figure}[htb]
\begin{center}
\mbox{\begin{tabular}[t]{cc}
\subfigure[]{\includegraphics[width=0.50\textwidth]{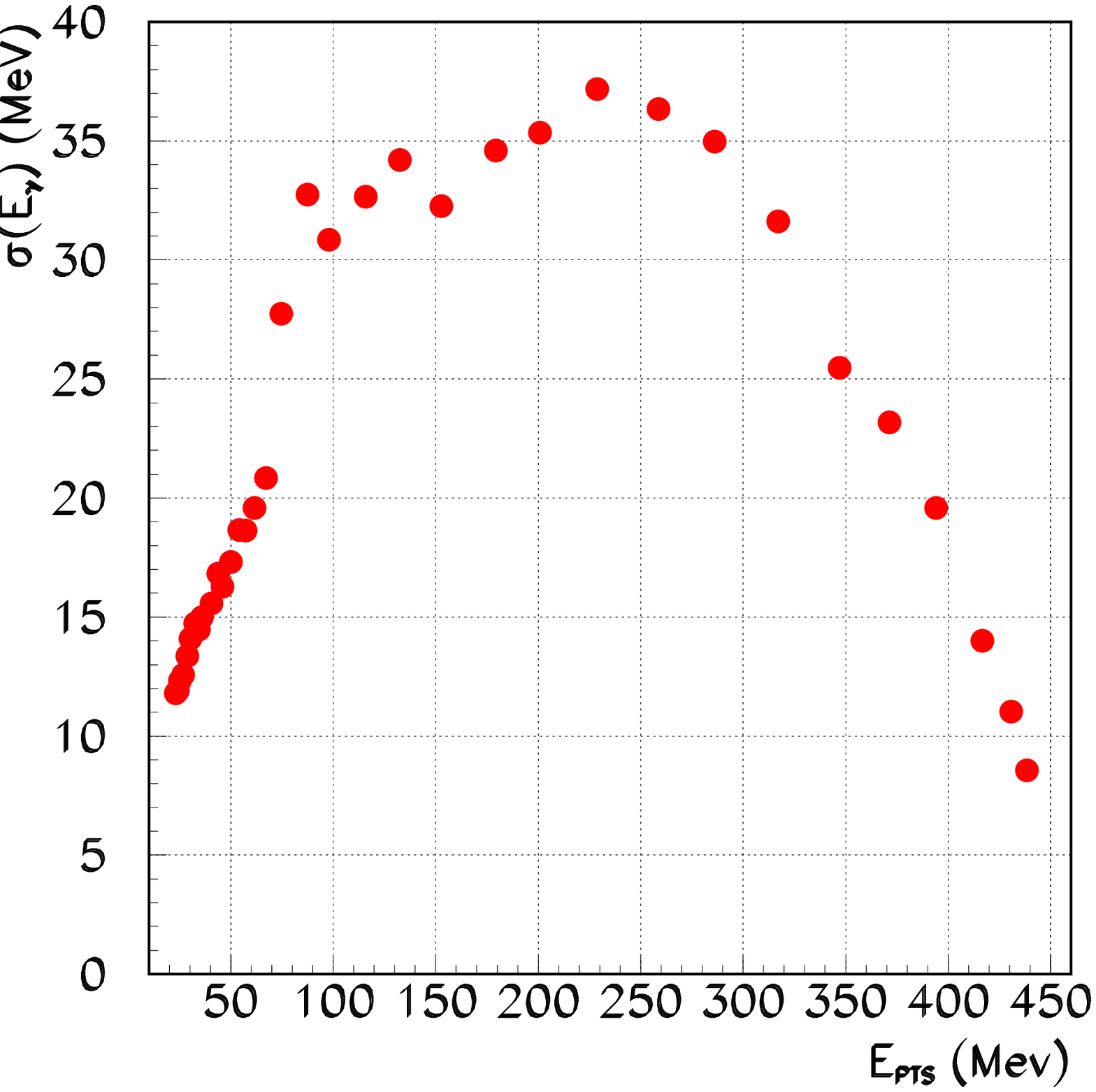}
\label{sigene}
} &
\subfigure[]{\includegraphics[width=0.50\textwidth]{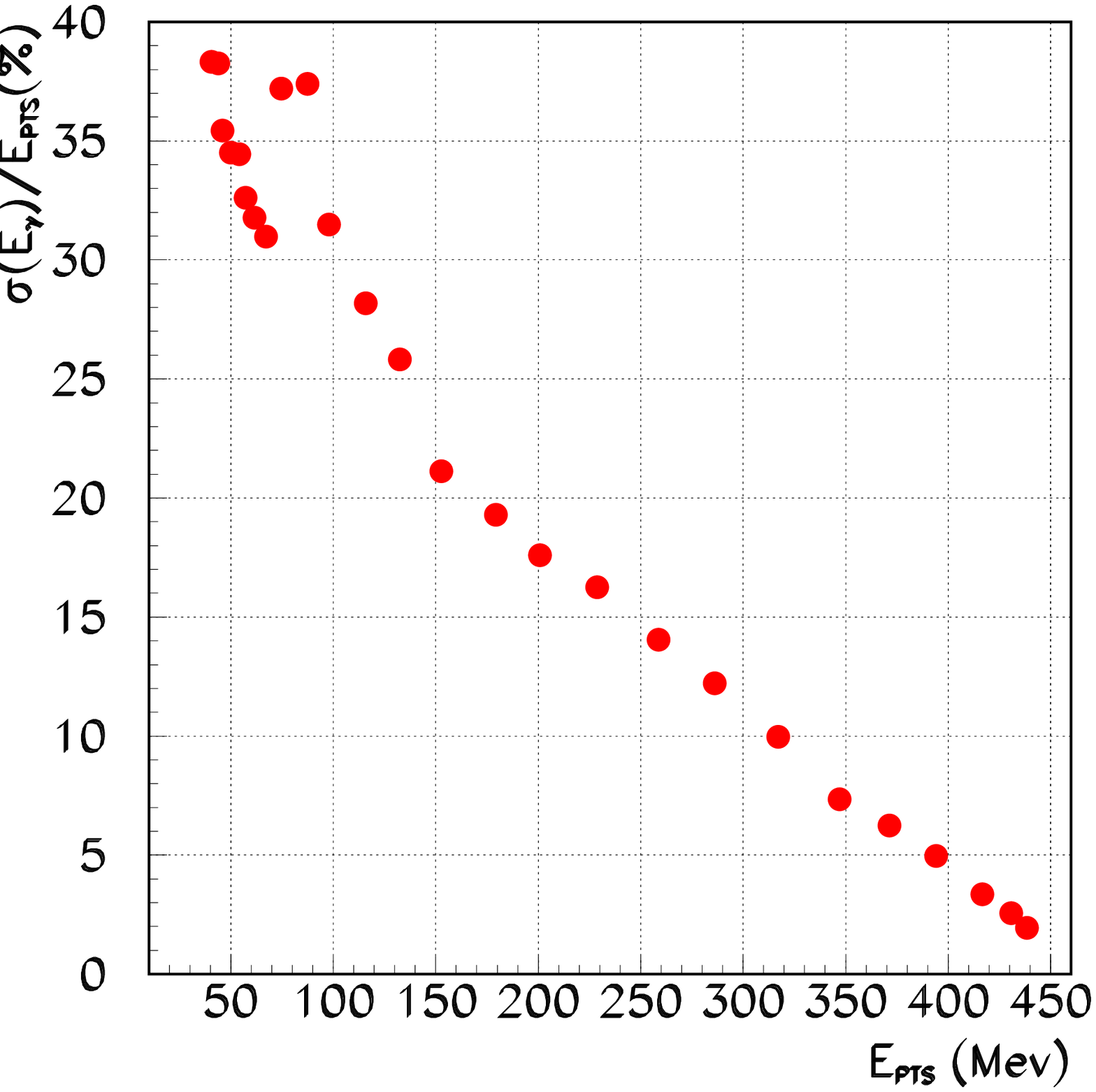}
\label{sigrelene}
} \\
\end{tabular}}
\end{center}
\caption{Absolute a) and relative b) spread in $E_\gamma$ for events with 3e$^-$ per 
bunch following Poisson distribution versus $E_{PTS}$.}
\end{figure}

\noindent
The absolute and relative RMS spreads in the $E_\gamma$ versus $E_{PTS}$ 
distribution limited to the central band, that is excluding outliers,
are shown in Fig.\ref{sigene}-\ref{sigrelene}. 
Outliers constitute at most a few percent of the events and their distribution
cannot be easily parameterized.

\section{Conclusions}

The BTF/PTS has been described 
in detail. It has been characterized by studying the data taken in LNF during 
the AGILE calibration campaign. Analysis of the target data has allowed to 
characterize the $e^-$ beam precisely.\\
The parameters of the Monte Carlo generator have been mostly determined from the data.
Corrections to the data have been necessary to account for a background
contamination not included in the simulation.\\
The relation between PTS coordinate and $\gamma$-ray energy has been 
calibrated with the Monte Carlo simulation and the $E_{PTS}$ distributions 
from simulation and data are in good agreement.\\
This calibration of the system allows to use it for calibrating photon 
detectors, like the ST of AGILE, as presented in a forthcoming paper.\\
Important parameters of the system like point spread function, effective area 
and energy resolution versus the $\gamma$ energy has been extracted through the Monte 
Carlo simulation validated with this system and used to extract relevant scientific 
results since the AGILE launch in 2007 \cite{agimis2}.

\section*{Appendix}

The photon flux can be predicted analytically with some approximated formulae 
from Quantum Electrodynamics to be compared with the Monte Carlo predictions.\\
The formula for the bremsstrahlung differential cross section for photon emission 
with energy $E_\gamma$ from an electron with energy $E_e$ is, with good approximation
\cite{pdb} 
\begin{equation}
\frac{d\sigma}{dy} = \frac{A}{X_0N_Ay} \left(\frac{4}{3} - \frac{4}{3}y + y^2\right)
\label{sigdy}
\end{equation}
where $A$ is the atomic number of the material, $X_0$ its radiation length, $N_A$ 
the Avogadro number and $y=E_\gamma/E_e$.\\
Eq.\ref{sigdy} can be integrated to predict the number of photons 
with relative energy between $y_{min}$ and $y_{max}$ 
emitted by a radiator of thickness $d$

\begin{equation}
 \overline N_\gamma(y_{min},y_{max}) = \frac{d}{X_0}\left[\frac{4}{3}\ln\left(\frac{y_{max}}{y_{min}}\right) - 
\frac{4}{3}(y_{max}-y_{min}) + \frac{1}{2}(y_{max}-y_{min})^2 \right]
\label{ngeq}
\end{equation}

\noindent
In our setup the radiator consists of four silicon layer $410\,\mu \mathrm{m}$ 
thick, a window of Be $0.5\,\mathrm{mm}$ thick and about $35.0\,\mathrm{cm}$ of Air.\\
The total thickness $d$ expressed in radiation lengths is 
(with $X_0^{Air} = 30500.0\,\mathrm{cm}$, $X_0^{Be} = 35.3\,\mathrm{cm}$ and 
$X_0^{Si} = 9.36\,\mathrm{cm}$)
\[
\frac{d}{X_0} \approx \frac{35.0\,\mathrm{cm}}{X_0^{Air}} + \frac{0.05\,\mathrm{cm}}{X_0^{Be}} + 
\frac{4\times 0.041\,\mathrm{cm}}{X_0^{Si}} \approx 2.0\,10^{-2}
\]

\noindent
$N_\gamma$ follow the Poisson distribution.
Also the number of electrons per bunch follows a Poisson distribution with average 
$\overline N_e$, therefore the probability of emitting 1 or more photons is
\begin{eqnarray}
P_{N_e}(N_\gamma=1) &\approx & \overline N_e \overline N_\gamma - (\overline N_e\overline N_\gamma)^2 \\
P_{N_e}(N_\gamma>1) &\approx & \frac{(N_eN_\gamma)^2}{2} 
\end{eqnarray}

\bibliographystyle{unsrt}
\bibliography{BTFPaper}

\end{document}